\documentclass[11pt,english]{article}

\usepackage{acorn}
\title{AI-Ready Control System for the Fermilab Accelerator Complex}
\TMNumber{FERMILAB-TM-2928-AD-DI-PIP2}

\begin{document}

\maketitle

\pagenumbering{roman}

    \setlength{\parindent}{0pt}
    \addcontentsline{toc}{section}{Acronyms}
\section*{Acronyms}

\begin{center}
{\footnotesize
    \begin{longtblr}[theme=blank,entry=none]{
    hlines,
    vlines,
    row{1} = {bg=acornBlue, fg=white, font=\bfseries},
    rowhead = 1,
    column{1} = {0.75in,l,m},
    column{2} = {5.35in,l,m},
    }
        Acronym & Meaning 
        \\ ACNET & Accelerator Network (Fermilab's legacy accelerator control system)
        \\ ACORN & Accelerator Controls Operations Research Network project
        \\ ADC & Analog to Digital Converter
        \\ AI/ML & Artificial intelligence / machine learning
        \\ AIP & Accelerator Improvement Plan (a type of funded project)
        \\ ASIC & Application-specific integrated circuit
        \\ BNB & Fermilab's Booster Neutrino Beamline
        \\ CAMAC & Computer Automated Measurement and Control module
        \\ CPU & Central processing unit
        \\ DIKW & ``Data, Information, Knowledge, Wisdom'' (DIKW) pyramid~\cite{Ackoff1989}
        \\ DOE & Department of Energy
        \\ Elog & Electronic logbook (used by Fermilab's accelerator control room)
        \\ EPICS & Experimental Physics and Industrial Control System
        \\ FPGA & Field programmable gate array
        \\ GPU & Graphics processing unit
        \\ HEP & High Energy Physics
        \\ IOC & Input/Output Controller (element of the EPICS control system)
        \\ KPP & Key Performance Parameter
        \\ L-CAPE & Linac--Condition Anomaly Prediction of Emergence
        \\ LBNF & Fermilab's Long Baseline Neutrino Facility beamline
        \\ LLM & Large Language Model
        \\ MI & Fermilab's Main Injector synchrotron
        \\ MLOps & Machine Learning Operations
        \\ Mu2e & Fermilab's muon to electron conversion experiment
        \\ NIST & National Institute of Standards and Technology
        \\ PID & Proportional-integral-derivative algorithm
        \\ PIP-II & Fermilab's Proton Improvement Plan II superconducting linac
        \\ PLC & Programmable Logic Controller
        \\ RR & Fermilab's Recycler storage ring
        \\ SoM & System-on-module circuit boards
        \\ SY & Fermilab's Switchyard external beamlines
        \\ TLG & Timeline Generator (sequence of accelerator actions)
        
    \end{longtblr}
}
\end{center}
    \setcounter{table}{0}

\tableofcontents
\listoftables
\listoffigures

\newpage
\pagenumbering{arabic}

\renewcommand{\labelenumii}{\arabic{enumi}.\arabic{enumii}}
\renewcommand{\labelenumiii}{\arabic{enumi}.\arabic{enumii}.\arabic{enumiii}}
\renewcommand{\labelenumiv}{\arabic{enumi}.\arabic{enumii}.\arabic{enumiii}.\arabic{enumiv}}

\section{Purpose}\label{sec:Purpose}
The Fermilab Accelerator Complex is the largest national user 
facility in the Department of Energy (DOE) High Energy Physics (HEP) 
program and the only national user facility operating at Fermilab. 
The entire complex is operated with a single accelerator control 
system that initiates particle beam production; controls beam energy 
and intensity; transports particle beams to research facilities; 
measures beam parameters; and monitors beam transport through the 
complex to ensure safe, reliable, and effective operations. Reliable, 
high-intensity operation of the complex is critical to the 
success of the laboratory's flagship program, the Long-Baseline 
Neutrino Facility and Deep Underground Neutrino Experiment 
(LBNF/DUNE). This 
document describes what is needed to make the accelerator control 
system \textbf{\textit{AI-ready}} to support the development, 
deployment, and routine use of AI/ML techniques that improve operational reliability and help 
meet the demands of the current and future HEP research program.

\subsection{Executive Summary}
Community workshops and stakeholder interviews with Fermilab accelerator scientists, 
engineers, and operators have outlined a vision 
for a modernized accelerator control system that supports routine use 
of AI/ML. Section~\ref{sec:Background_and_Motivation} provides background on 
the current control system, the role of automation, and how AI/ML 
fits into the vision for enhanced operations. To prepare Fermilab's 
accelerator control and monitoring infrastructure to robustly support 
AI/ML integrations, the following three systems are recommended:

\begin{enumerate}
    \item Machine learning operations framework (Section~\ref{sec:Lifecycle_Management_of_Automations}): tools and processes to develop, deploy, and keep track of AI/ML‑based automation in a reliable and reproducible way.
    \item Data quality prerequisite and framework (Section~\ref{sec:Data_quality_and_access}): defines what ``good enough for AI/ML'' means for accelerator data. For example, that it is complete, self-consistent, timely, secure, accurate, standardized, and not systematically biased.
    \item Workflow integration with LLMs 
(Section~\ref{sec:Workflow_integration_with_llms}): enables 
accelerator physicists, engineers, developers, and operators to find 
information, write and review code, and streamline routine 
communication and analysis tasks, in a secure way.
\end{enumerate}

Building on the vision articulated through community input, these recommendations provide a practical path forward. Supporting use cases that illustrate how these systems apply across controls, diagnostics, and support systems are detailed in Section~\ref{sec:AI_Prerequisites_and_Use-Cases}. Together, they underscore that high-quality, standardized data is the foundation for making any of this work.

\section{Background and Motivation}\label{sec:Background_and_Motivation}
\subsection{Document Project Context}
This document was developed to guide the Accelerator Controls 
Operations Research Network (ACORN) project at Fermilab with the mission need to modernize the accelerator 
control system by enabling AI/ML. It is based on community workshops 
and stakeholder interviews, and outlines what is necessary to meet 
objective key performance parameters established for the 
project's ``Controls Infrastructure and Applications'' team:
\begin{displayquote}
\textit{For the [team's] objective key performance parameter, the project will implement the 
cybersecurity enhancements and implement support for the development, 
deployment, and monitoring of AI/ML capabilities to optimize 
accelerator operations, predict failure modes, and provide new 
simulation capabilities.}
\end{displayquote}
While written for ACORN, the frameworks and practices described here 
may be useful to other accelerator facilities facing similar 
challenges in making their control systems AI-ready.

\subsection{Fermilab Accelerator Complex}

The Fermilab Accelerator Complex uses a linear accelerator (Linac), 
the Booster rapid-cycling synchrotron, the Main Injector synchrotron, 
and the Recycler storage ring to produce two primary proton beams: an 
8~GeV beam from the Booster and a 120~GeV beam from the Main 
Injector. The current Linac will be replaced by the PIP-II 
superconducting linear accelerator, which will provide higher 
intensity beams to support megawatt-class operations for LBNF/DUNE. 
The research program uses these proton beams to produce muons and 
neutrinos at world-leading intensities. Beam transfer lines transport 
particle beams from the accelerators to the experiments executing 
HEP's Particle Physics Research Program. 

\begin{figure}[ht]
    \centering
    \includegraphics[width=0.8\linewidth]{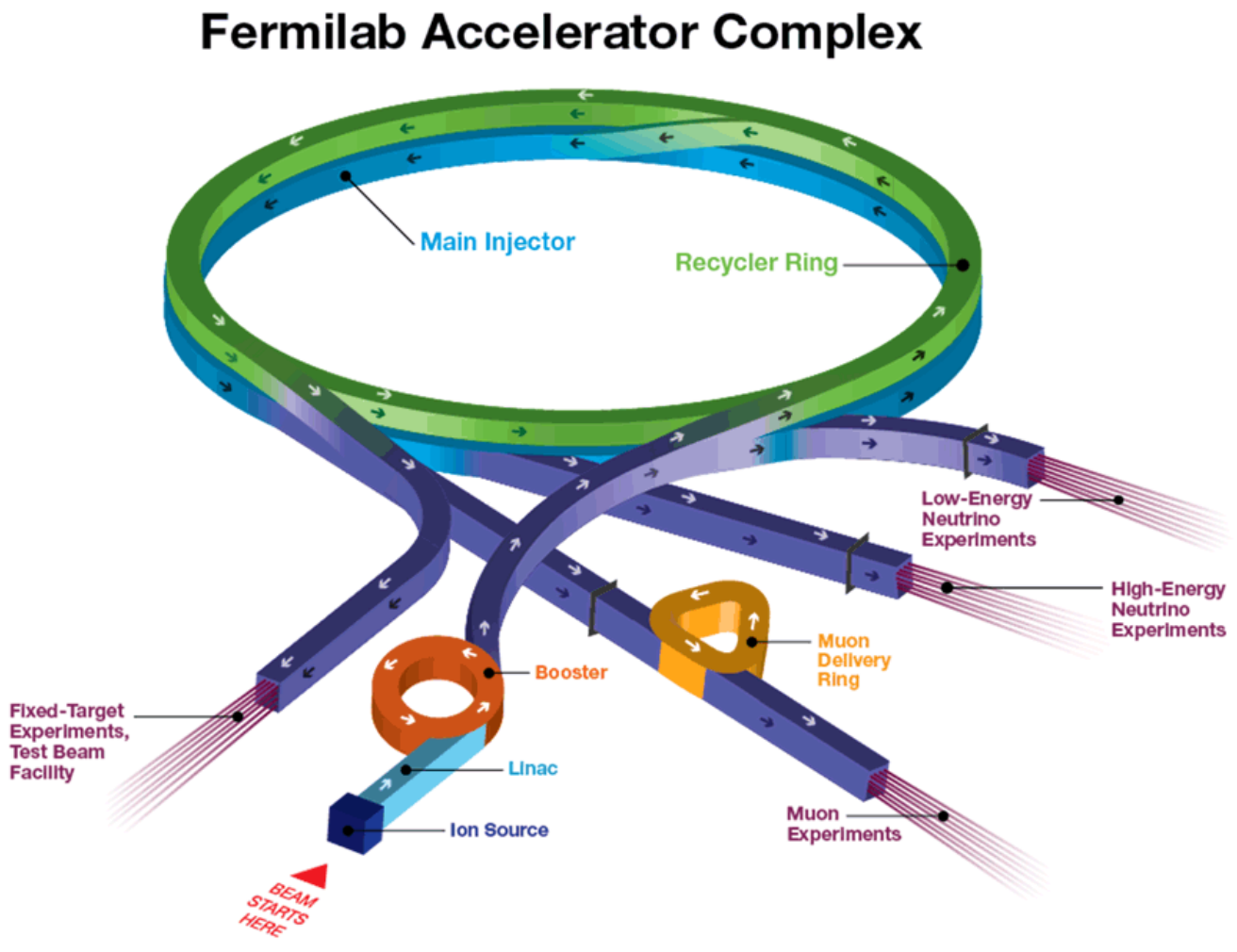}
    \caption{Layout of the current Fermilab Accelerator Complex showing the 
major accelerators and beam transfer lines.}
    \label{fig:FAC}
\end{figure}

Figure~\ref{fig:FAC} shows the layout of the accelerator 
complex and its major components. The entire complex is managed by a single control system, as 
introduced in Section~\ref{sec:Purpose}. The following subsections 
provide background on how this control system works today, the role 
of automation, and how AI/ML fits into the picture.

\subsection{Control System}
Accelerator systems can be regulated manually, automatically, or by a 
combination of the two. \mbox{Fermilab's} accelerator complex uses a mixture 
of all three approaches.

\begin{figure}[ht]
    \centering
    \includegraphics[width=1\linewidth]{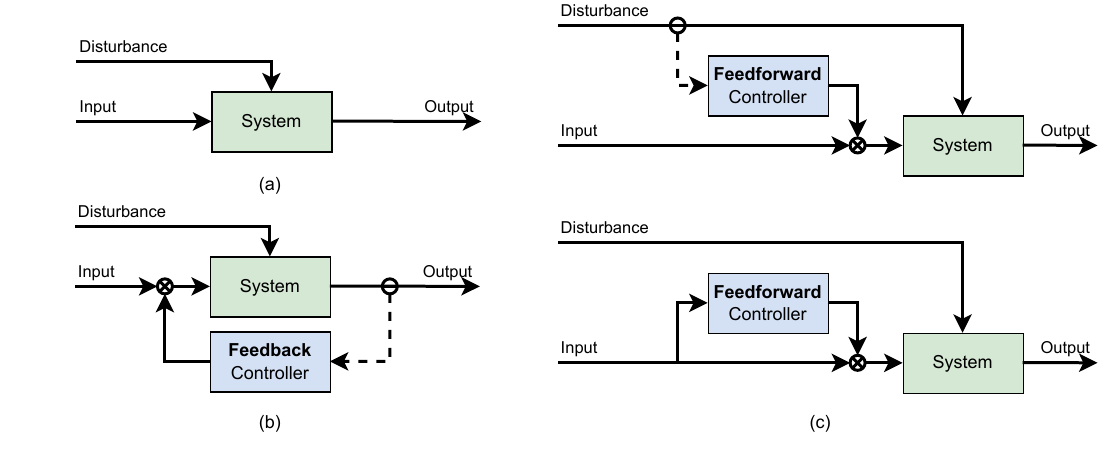}
    \caption{(a) Manual or open-loop controller, (b) feedback controller, and (c) two examples of feedforward controllers.}
    \label{fig:FF_and_FB}
\end{figure}

In manual (open-loop) control, an operator adjusts the system based on 
experience and observation, with no automatic correction, 
Figure~\ref{fig:FF_and_FB}(a). Prior to the advent of automatic 
control, all machines were run this way.

Feedback control adds sensors that measure the system's current state 
and automatically apply corrections to keep it close to the desired 
state, Figure~\ref{fig:FF_and_FB}(b). The most common example is the 
Proportional-Integral-Derivative (PID) controller, which responds to 
current error, accumulated past error, and predicted future error.

Feedforward control goes a step further, using mathematical models to 
anticipate disturbances and act before errors occur, 
Figure~\ref{fig:FF_and_FB}(c). An example is model predictive control, 
which computes the best current inputs based on where the system is 
expected to be in the future.

Given the complexity of the machines at Fermilab, there is ample 
opportunity to implement more automatic controllers. These can range 
from simple single-input, single-output systems using PID 
algorithms to more complex multi-input, multi-output systems 
that employ trained machine learning (ML) models. Depending on 
performance requirements, these controllers can run centrally on 
servers or be deployed on dedicated field hardware closer to the 
machine.

The ``accelerator control system''\footnote{Referred to as the 
``control system'' in this document.} orchestrates these controllers 
to manage the interconnected components that produce, accelerate, and 
deliver particle beams. While much of the control system is already 
automated, there are still many higher-level actions that can and 
should be automated.

\subsection{System Automation}
The purpose of system automation is to achieve reliable performance that meets defined objectives while maintaining robustness.
The controllers in the accelerator’s control system adapt quickly and reliably to changing conditions and disturbances, as long as these remain within an expert-defined range.
Often, machine conditions drift outside the range of a deployed controller, and an expert must periodically come to the control room to perform beam scans and measurements and derive and apply updated settings for the controllers.

Automating the process of performing scans, analyzing measurements, 
and updating controller settings would enable faster responses to 
common operational drifts and reduce the burden of repetitive expert 
interventions. Automation levels can be characterized using the NIST 
``Autonomy Levels for Unmanned Systems (ALFUS) Framework,''~\cite{Huang2008} which 
defines five levels based on how much a system can accomplish without 
human involvement.

\begin{figure}[ht]
    \centering
    \Large{\textbf{ALFUS System Autonomy Levels}}
    \includegraphics[width=1\linewidth]{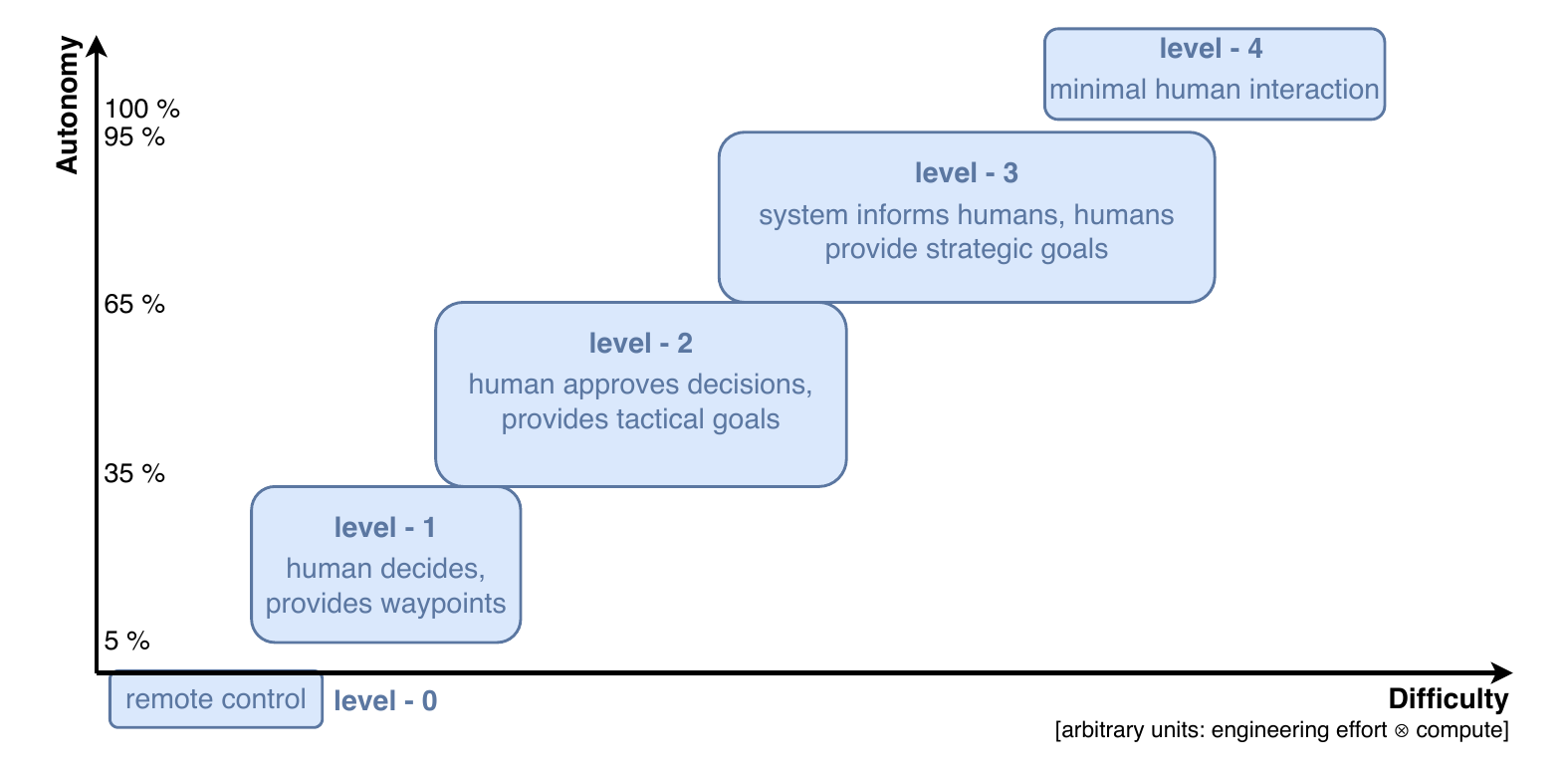}
    \caption{The ``Autonomy Levels for Unmanned Systems (ALFUS) Framework'' defines five autonomy levels based on the required amount of human interaction to achieve a system's mission. Level~0 requires constant human involvement, similar to how a remote control depends on a human to push its buttons. At the other end of the spectrum, Level~4 represents a system capable of functioning with minimal human interaction.}
    \label{fig:autonomy_levels}
\end{figure}

Currently, most automations at Fermilab's accelerator complex operate 
at Level~1 or below. Level~0 automations consist of teleoperations 
and direct parameter changes in ACNET (e.g., manual magnet current 
settings). Level~1 includes autotune scripts that specify target 
values, PID loop controller configurations, and deterministic beam scan scripts. 
Level~2 automations involve beam scans with online optimization, 
though these have largely remained in the development stage under the 
supervision of system experts.

\begin{figure}[ht]
    \centering
    \Large{\textbf{DIKW Pyramid and Decision-Making Complexity}}
    \includegraphics[width=1\linewidth]{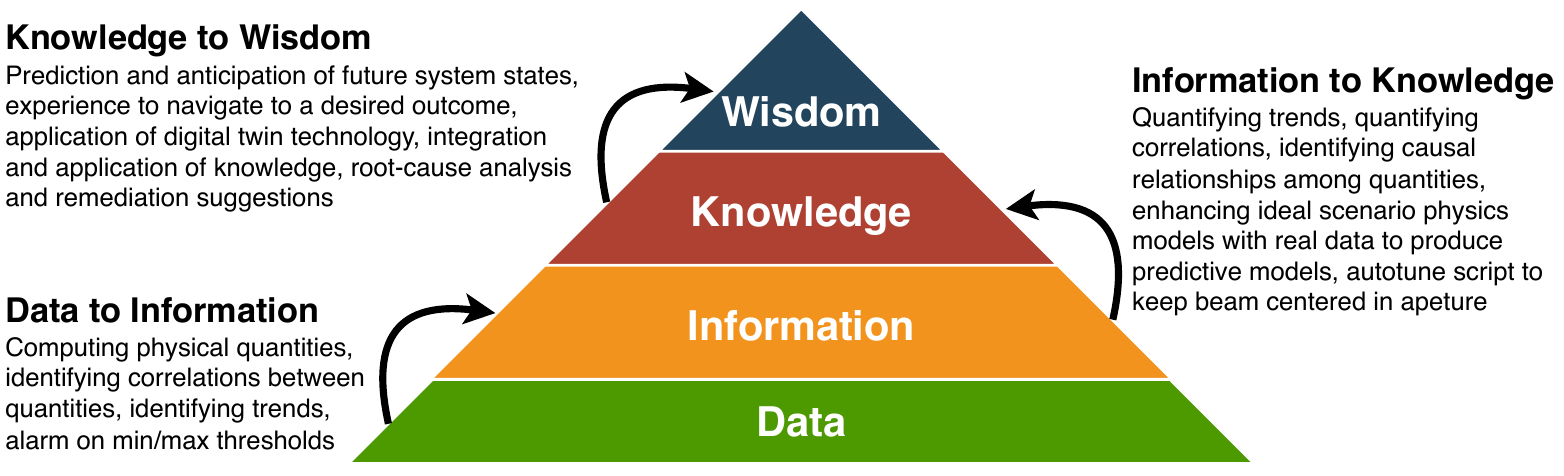}
    \caption{A system must transmute data into information, information into knowledge, and add experience and situational awareness to create the wisdom necessary to make trustworthy decisions.}
    \label{fig:integration_complexity_levels}
\end{figure}

A useful way to think about the complexity of these automations is 
Ackoff's ``Data, Information, Knowledge, Wisdom'' (DIKW) pyramid~\cite{Ackoff1989} 
(Figure~\ref{fig:integration_complexity_levels}), which describes how 
raw data must be progressively refined before it can support 
trustworthy decisions. Most current automations operate at the data 
and information levels, while higher-level decisions continue to rely 
heavily on operator and machine expert knowledge and wisdom. Advancing 
higher autonomy levels will require machine experts to formalize 
their operational practices so they can be codified into automated 
methods that operate at the knowledge and wisdom levels.

\subsection{Relationship of Automations and Current Accelerator Control System}

Fermilab's control system currently manages over 500,000 devices, 
setpoints, and readbacks, with 30,000 accessible only through CAMAC 
infrastructure established over 50 years ago. The most common 
automations today are control-loop automations, including machine 
protection systems, beam instrumentation integration and calculation, 
sequencer operations, Fast-Time Plotter, PLCs, regulation scripts, 
and damper systems. These have evolved over decades of commissioning 
and operational experience to standardize routine operations and 
accelerate decision-making.

To achieve higher autonomy levels, machine experts need reliable data 
with which to build automated algorithms for decision-making. A key 
objective of the ACORN project is to provide machine experts with the 
data and streamlined deployment patterns they require. However, 
systems built primarily to support near-term beam delivery are not 
always well suited for the data pathways that AI/ML applications 
require. ACORN presents an opportunity to address this gap through 
technical innovation and organizational collaboration.

\subsection{AI/ML}

Artificial Intelligence and Machine Learning (AI/ML) represent a broad class of computational approaches for pattern recognition, prediction, and decision-making in complex systems. Rather than relying solely on fixed, explicitly programmed rules, AI/ML methods learn from data to model complex relationships, adapt to changing conditions, and generalize from past observations. This makes them well suited to domains where many coupled variables, non-linear dynamics, and incomplete models limit the effectiveness of traditional control and analysis techniques.

\begin{figure}[ht]
    \centering
    \Large{\textbf{Methodologies for Unsupervised Anomaly Detection}}
    \includegraphics[width=0.9\linewidth]{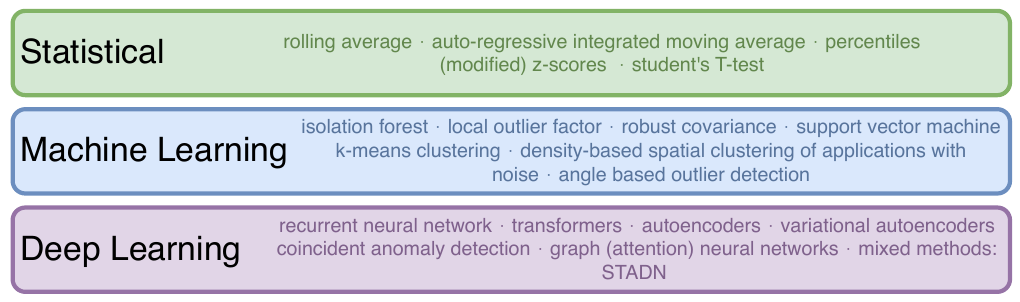}
    \caption{Methodologies for unsupervised anomaly detection. Presently the most rudimentary method is primarily used: setting warn and alarm thresholds on parameters.}
    \label{fig:Anomaly_detection_methods}   
\end{figure}

In accelerator operations, AI/ML technologies can identify anomalies, predict equipment failures, create virtual diagnostics, and increasingly optimize entire accelerator systems. They can detect anomalous behavior ranging from simple threshold violations and trending drifts to complex patterns involving correlations across multiple signals. Detection methodologies span statistical techniques such as rolling averages to machine learning and deep learning approaches capable of identifying correlated shifts across many variables at once (Figure~\ref{fig:Anomaly_detection_methods}, adapted from~\cite{Iliopoulos2023}). Similarly, AI/ML-enabled forecasting extends traditional simulation-based approaches by handling systems that are too complex or time-intensive for pure simulation alone, using models that combine physics-based simulations with data-driven components trained on operational data.

The implementation of AI/ML systems requires robust infrastructure 
including: high-quality data with consistent timestamp policies and 
efficient data pathways (see 
Section~\ref{sec:Data_quality_and_access}); and comprehensive MLOps capabilities, including reliable mechanisms for 
data and model version tracking and model monitoring and maintenance automation 
(see Section~\ref{sec:Lifecycle_Management_of_Automations}).

\subsection{Vision for an AI/ML-Enhanced Accelerator Control System}

The vision for a modernized accelerator control system has been shaped 
through community workshops and stakeholder interviews with 
accelerator experts at Fermilab. The modernized system will integrate 
traditional control methods with AI/ML capabilities to create a more 
adaptive and efficient operational environment. This integration will 
help transform raw accelerator data into actionable knowledge, 
allowing operators and machine experts to build more robust 
decision-making processes and move toward higher autonomy levels.

Advanced AI/ML diagnostics and controls will create new interfaces 
that link accelerator instrumentation with the control system and 
present actionable insights to operators and machine experts. These 
capabilities will support complex-wide optimization objectives, 
including maximizing up-time, minimizing beam losses, reducing power 
consumption, and making more efficient use of human expertise.

\begin{figure}[ht]
    \centering \includegraphics[width=1\linewidth]{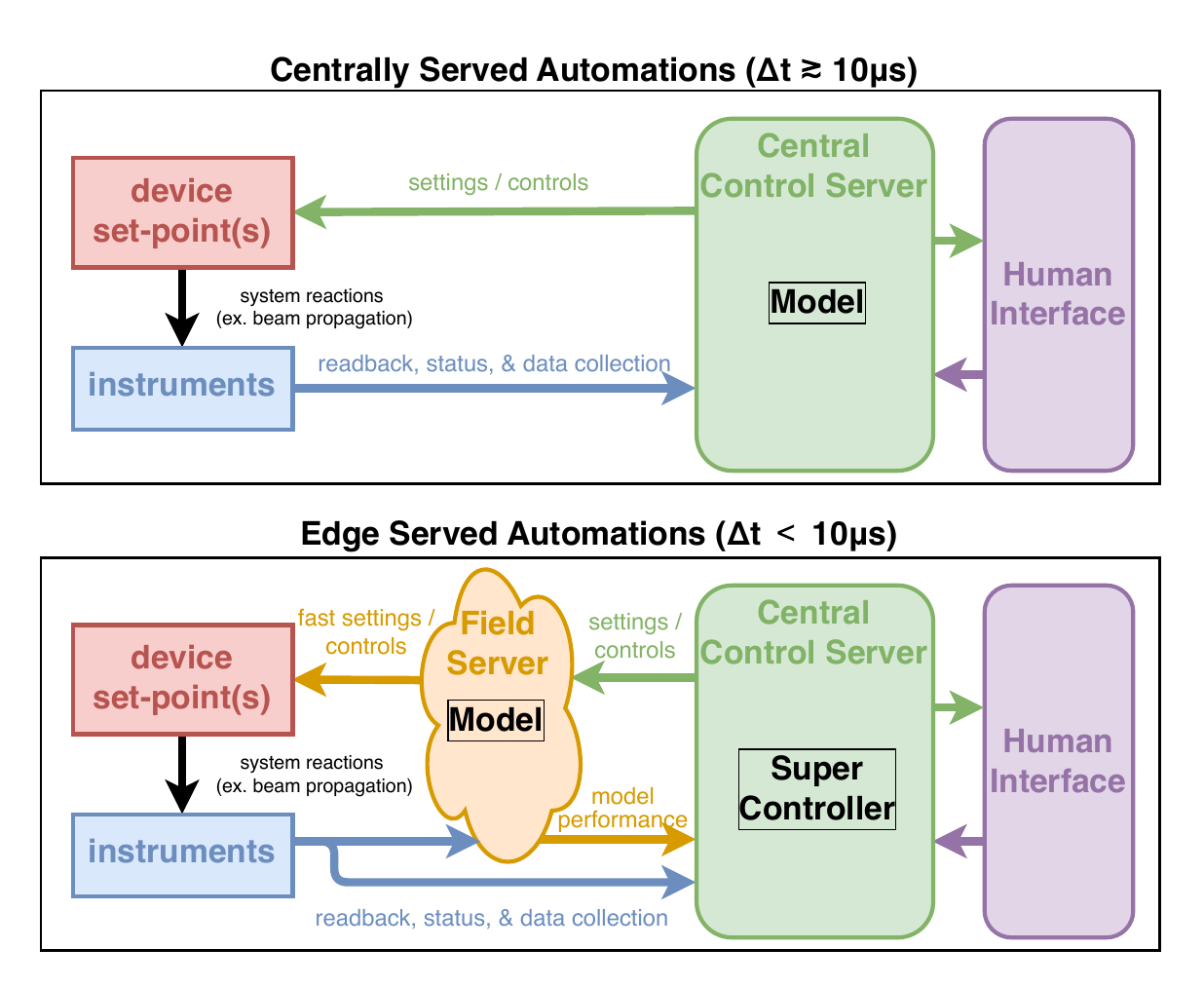}
  \caption{Centrally served automations ($\Delta t \geq 10\,\mu s$) support multi-system optimization, while edge served automations ($\Delta t < 10\,\mu s$) support low-latency control loops.}
    \label{fig:cartoon_controls-Combined_3}
\end{figure}

\begin{figure}[ht]
    \centering \includegraphics[width=0.85\linewidth]{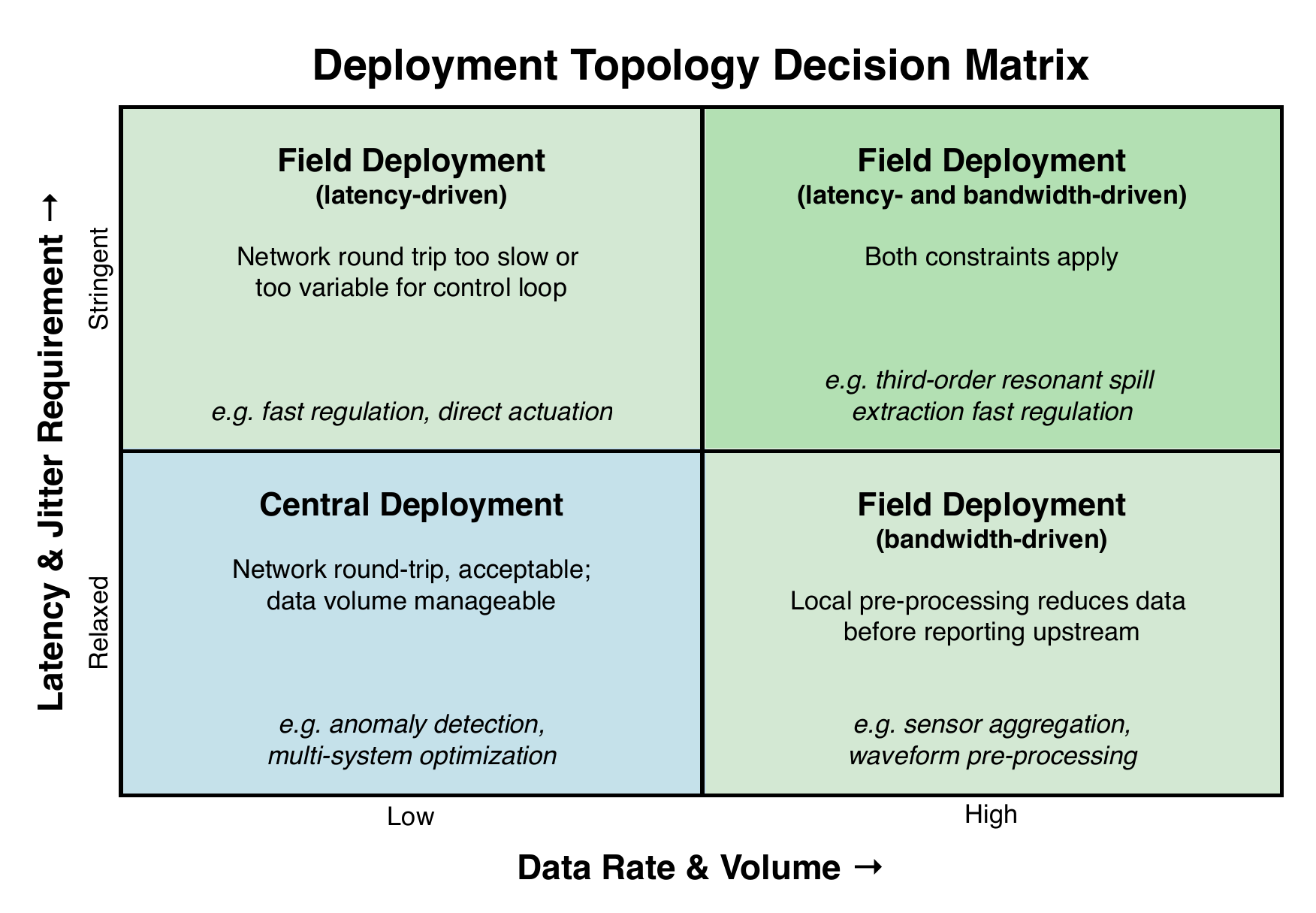}
  \caption{Deployment choice depends on the control-loop latency (time to read a signal, compute an action, and deliver that action) and jitter, and on data rate and volume.}
    \label{fig:Deployment_Topology_Matrix}
\end{figure}

ACORN will support two approaches to deploying AI/ML.
The architecture for each is shown in Figure~\ref{fig:cartoon_controls-Combined_3}.
Centrally deployed automations are suited for cases where control-loop response times and data volumes are not limiting constraints, supporting complex, multi-system optimization.
Field-deployed automations are appropriate when one or more of the following conditions apply: the control loop requires a response faster than the facility network can reliably deliver; the volume or rate of data generated at the source makes continuous central transmission impractical; or both constraints apply simultaneously, requiring local computation that is both low-latency and bandwidth-reducing.
In all field-deployment cases, aggregate performance metrics and status information are reported back to the central control server to support monitoring and governance.
These deployment conditions are summarized in Figure~\ref{fig:Deployment_Topology_Matrix}.


Key applications include root-cause analysis, optimization across 
multiple variables and objectives, active learning controls, digital 
twin development, and predictive maintenance. For example, rather 
than pursuing only simple control objectives such as ``maintain 
magnet current,'' the system will be able to optimize for 
higher-level goals like ``maintain low losses in the Booster while 
maximizing beam power,'' which necessarily includes keeping magnet 
currents where they need to be. This represents a fundamental shift 
from reactive to predictive operations, in which the system 
anticipates and prevents problems rather than merely responding to 
them.

The enhanced control system will preserve the reliability and safety 
of traditional approaches while adding the adaptability of AI/ML 
methods, providing a strong basis for future accelerator operations 
and scientific discovery.

This vision emerged machine-by-machine and system-by-system during 
formal requirements interviews with experts across 
departments. A consolidation of these interviews is presented in 
Section~\ref{sec:AI_Prerequisites_and_Use-Cases}. While ACORN cannot 
deliver every automation envisioned, it can redesign the control 
system with this vision as a guiding framework. Just as HEP 
experimental detectors have advanced over decades to expand discovery 
potential and precision measurements, ACORN has the opportunity to 
modernize the accelerator control system by ensuring that machine 
experts have the data, data quality, and data pathways needed to 
apply AI/ML methods that enhance the reliability and performance of 
the U.S.'s flagship accelerator complex.

\section{Lifecycle Management of Automations: MLOps} \label{sec:Lifecycle_Management_of_Automations}
Machine learning operations (MLOps) refers to the end-to-end process 
of researching, developing, testing, deploying, monitoring, and 
improving ML models in operational systems. At their core, ML models 
are algorithms that learn patterns from example data (called training 
data) and use those patterns to generate predictions or responses 
from new input. During training, the model adjusts internal 
parameters, known as ``weights,'' until its predictions match the 
training data well enough. An MLOps framework is a defined set of 
tools and practices that standardize and automate each step of this 
process, ensuring that deployed models are reproducible, validated, 
and observable.

A fundamental prerequisite for building ML models is access to 
high-quality data along with knowledge of the accelerator, how it 
functions, and its limitations. How control-system data is collected 
and made available is detailed in Section~\ref{sec:data_access} 
(Data Access). The following subsections detail each step of the 
MLOps process. The final subsection describes the overall 
observability and human interface to the MLOps pipeline 
itself\footnote{The human interface for overseeing the 
\textit{MLOps pipeline} is distinct from the human interface for 
overseeing \textit{models hosted by MLOps}.}.

\begin{figure}[ht]
    \centering
    \includegraphics[width=1\linewidth]{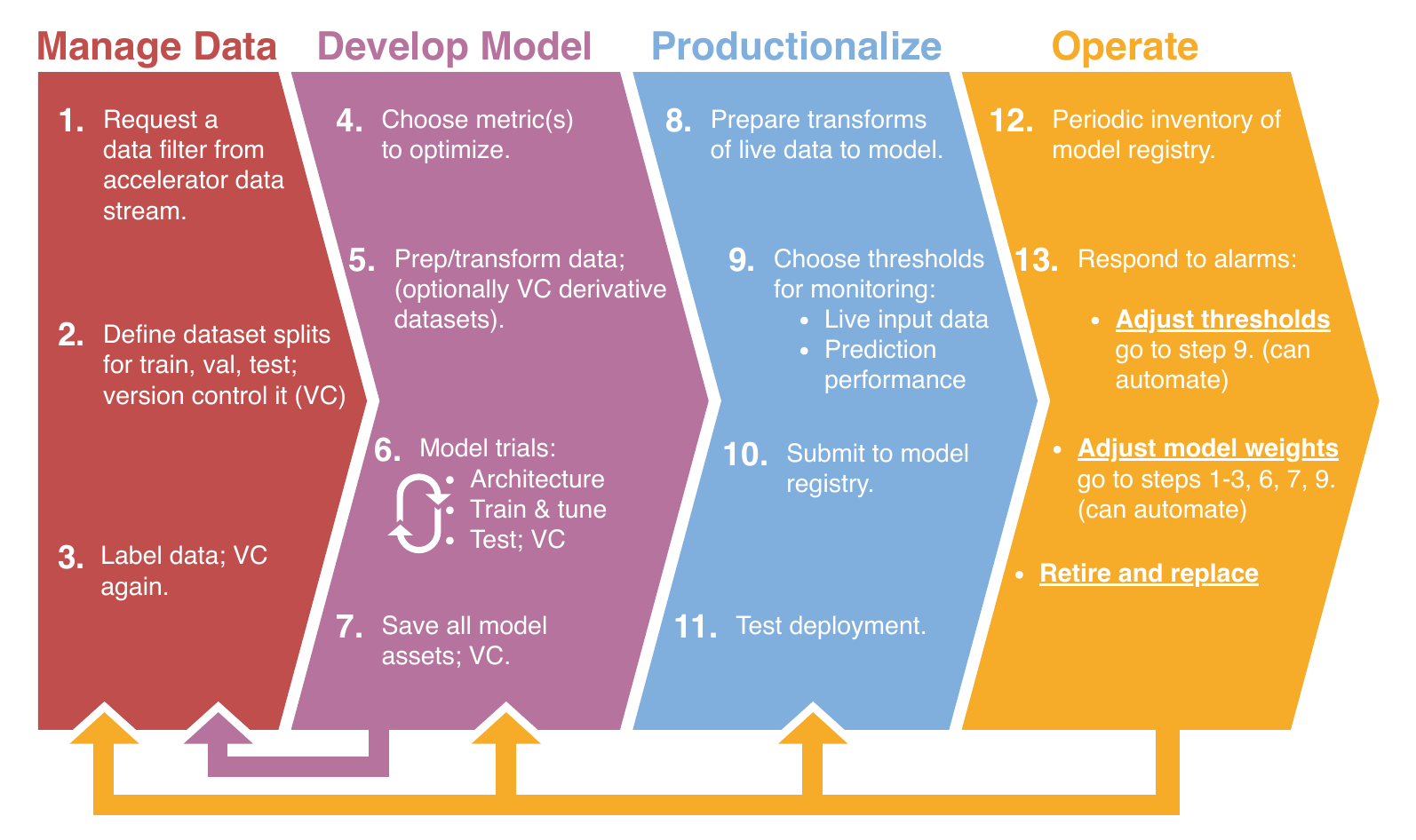}
    \caption{Lifecycle phases of AI/ML solution development and deployment (MLOps). Arrows pointing back to previous steps may be triggered manually or automatically as maintenance of AI/ML models themselves become more autonomous. Data management and model development phases apply equally to central and edge deployments. Productionalization and operation differ for edge deployments, requiring additional model optimization, hardware-specific deployment procedures, and a longer retrain-to-redeploy cycle; details are in Sections~\ref{sec:mlops:productionalize} and~\ref{sec:mlops:operate}..}
    \label{fig:MLOPs-Engineer-Arrows}
\end{figure}

Figure~\ref{fig:MLOPs-Engineer-Arrows} provides an overview of the 
MLOps pipeline, organized by information flow and broken into four 
main phases.

The first phase is ``Manage Data,'' which includes 1.~getting the 
data, 2.~exploring the data to define subsets, and 3.~labeling the 
data for model development. Defining appropriate data subsets and 
applying accurate labels are critical steps that are often 
underestimated. Equally important is retaining and versioning all 
training datasets for each model. This makes it possible to track 
how the data changes over time (known as data drift), which helps 
determine whether a model's performance has degraded and whether it 
needs to be retrained or replaced.

The second phase is ``Develop Model.'' This includes 4.~choosing a 
metric to optimize (i.e., defining what success looks like), 
5.~transforming or preparing the data for use by the model, 
6.~creating a baseline model and iteratively improving it, and 
7.~saving the model and its weights.

Phase~3 is ``Productionalize,'' in which the model is prepared for 
use in the accelerator control system. This stage is often the most 
challenging because it requires close collaboration between system 
experts (machine experts, controls and instrumentation engineers) and 
ML developers. Working together, they will 8.~define how live data is 
ingested and formatted for the model, 9.~define metrics for 
monitoring to ensure the model is behaving as expected, 10.~package 
everything for deployment and register it in a model registry, and 
11.~pass tests for all components.

The final phase is ``Operate,'' in which models are actively deployed 
in the control system. Depending on its goal, a model's output may 
identify anomalies, regulate subsystems, optimize systems, or predict 
needed maintenance. In Step~12, deployed models are periodically 
inventoried to ensure they remain relevant and correct. Step~13 
covers ongoing model maintenance: humans observe model behavior and 
respond to alarms that indicate the model's performance or its input 
data has changed in unexpected ways. Such changes may require 
adjusting thresholds, retraining the model, or replacing it. Over 
time, these maintenance activities may themselves be automated, as 
represented by the backward arrows in 
Figure~\ref{fig:MLOPs-Engineer-Arrows}.

Throughout the lifecycle of an AI/ML model, different expertise is 
needed at each phase. ACORN aims to provide an end-to-end example 
with templates for each step to guide future developers. In 
phases~1 and~2, machine and instrumentation experts will need to work 
together with an AI/ML developer to define the problem, select the 
right data, and evaluate candidate models. In phase~3, the AI/ML 
developer and machine expert will need to work with backend and 
frontend software developers to prepare the model for deployment. In 
phase~4, the AI/ML developer will be the steward of the model and 
ensure operators understand its behavior and how to respond to 
alarms.

\subsection{Data Management}
Data management for MLOps involves streamlining how data is 
collected\footnote{Prerequisites for data generation and formatting 
are described separately in 
Section~\ref{sec:Data_quality_and_access}.}, documented, processed, 
archived, and shared.

\subsubsection{Data Management Steps}
The following steps outline how users and developers manage data 
within the MLOps framework.

\begin{enumerate}
    \item \myuline{Select a problem}: Problems well-suited for ML 
    include those where complex patterns in existing data can be 
    learned and applied to new data for classification, regression, 
    or prediction. Many problems, however, may be addressed with 
    simpler approaches that do not require ML, such as regulation 
    scripts and statistical monitoring. For both ML and non-ML 
    problems, dataset management tools help maintain an organized 
    historical record of the data used during development.

    \item \myuline{Select data}: Choose observed and/or simulated 
    data to develop a model. This data should be saved into a 
    dataset volume.

    \item \myuline{Enter datasets into registry}: To track and share 
    datasets, create an entry in the metadata registry documenting 
    how each dataset was created.

    \item \myuline{Track derivative datasets in registry}: Split and 
    tag the data into ``train,'' ``validate,'' and ``test'' subsets. 
    Create derivative datasets with transformed variables (using 
    version-controlled code). Create derivative datasets with labels, 
    generated either manually or automatically, documenting the 
    method used in each case.

    \item \myuline{Archive datasets}: Archive datasets in a 
    structured repository to ensure reproducibility and long-term 
    accessibility for future development and validation.
\end{enumerate}

\subsubsection{Data Management Tooling}
The following tools are needed to support data management for safe 
AI/ML operations. Each tool is described in the following tables which describe its core functionality, user tools, administrative tools, and integration interfaces.

\begin{itemize}
    \item \textbf{Control System Data Interface}, see Section~\ref{sec:Data_quality_and_access}, Table~\ref{tab:tooling_control_system_data_interface}. This is used to \myuline{select data} from the control system devices to be saved into a dataset volume for \myuline{archiving datasets}. 
    \item \textbf{Instrumentation Data Interface}, see Section~\ref{sec:Data_quality_and_access}, Table~\ref{tab:tooling_instrumentation_data_interface}. This is used to \myuline{select data} from the intsrumentation devices to be saved into a dataset volume for \myuline{archiving datasets}. 
    \item \textbf{Dataset Storage Volume}, see Table~\ref{tab:tooling_dataset_storage_volume}. This is where data is stored from \myuline{selected data} both from the accelerator and from simulations.
    \item \textbf{Metadata Catalog}, seeTable~\ref{tab:tooling_metadata_catalog}. This is the \myuline{dataset registry} tool for \myuline{archiving datasets} to keep track of them and how they were produced.
\end{itemize}

\begin{longtblr}[
        caption={Tooling specification for \textbf{Dataset Storage Volume}.}, 
        label={tab:tooling_dataset_storage_volume}
    ]{
        hlines, vlines,
        cells = {valign=m,font=\footnotesize},
        column{1} = {bg=acornBlue, fg=white},
        column{1} = {0.85in,l,m},
        column{2} = {5.15in,l,m}
    }
\textbf{Core Functionality} & Database to save datasets produced by queries to the control system, instrumentation, derivative datasets (human-labeled data records, subsets of data, etc.), and by simulation codes. \\ 
\textbf{User Tools} & Write permissions.\\
\textbf{Administrative Tools} & Limit delete permissions.\\
\textbf{Integrations} & Accepts data from \textbf{Control System Data Interface}, \textbf{Instrumentation Data Interface}, and user development servers and users. Readable by the \textbf{Metadata Catalog}, user development servers, and users. Allows integrations for retention policy automations.
\end{longtblr}

\begin{longtblr}[
        caption={Tooling specification for \textbf{Metadata Catalog}.}, 
        label={tab:tooling_metadata_catalog}
    ]{
        hlines, vlines,
        cells = {valign=m,font=\footnotesize},
        column{1} = {bg=acornBlue, fg=white},
        column{1} = {0.85in,l,m},
        column{2} = {5.15in,l,m}
    }
\textbf{Core Functionality} & Stores metadata about datasets: immutable information [e.g. unique identifier for the dataset, location of data, progeny of the data (how it was made), parent datasets, manifesto of the dataset (names and formats of columns for example), number of entries of dataset, storage size of dataset], and mutable (but logged) information [e.g. owner, stakeholders, retention policy, permission groups for data access, curated project identifier tags, internal project tags (determined by owner or permission group)]. User interfaces: web-based graphical and command line.\\ 
\textbf{User Tools} & User interfaces: web-based graphical and command line. Ability to search, browse, and submit datasets. Ability to update mutable fields.\\
\textbf{Administrative Tools} & Integrations for retention policy automations, role-based user management, and administratively defined dataset entry fields and mutability. Secure (preferably SSO integration).\\
\textbf{Integrations} & Read and compute summary statistics from datasets in \textbf{Dataset Storage Volume}.
\end{longtblr}

\subsubsection{Data Management Risk Mitigation}
These data management tools mitigate accessibility and stewardship 
risks such as those in Table~\ref{tab:mlops_data_management_system_risk_mitigations}.

\begin{longtblr}[ 
        caption={\textbf{Data Management System} risks and mitigations.}, 
        label={tab:mlops_data_management_system_risk_mitigations}
    ]{
        hlines, vlines,
        row{1} = {bg=acornBlue, fg=white}, 
        rowhead = 1,
        cells = {valign=m,font=\footnotesize},
        column{1} = {2.5in,l},
        column{2} = {3.5in,l}
    }
\textbf{Risk} & \textbf{Mitigation}\\*
Duplication of datasets. & \SetCell[r=2]{l}{Common and accessible \textbf{Dataset Storage Volume} allow all interested users to read the data they need.}\\*
Shareability and re-use of datasets. & \\*
Overwriting / mutation of a dataset with the same name. & Administratively controlled mutation permission on \textbf{Dataset Storage Volume} and \textbf{Metadata Catalog}. \\*
Lost history / provenance of a dataset. & Maintain provenance through \textbf{Metadata Catalog}. \\*
Retention of datasets for operational models for benchmarking new model versions. & \SetCell[r=2]{l}{Maintain important datasets and metadata, and staged removal of obsolete data through data-retention policy for the \textbf{Dataset Storage Volume} and \textbf{Metadata Catalog}.}\\*
Uninhibited growth of storage. &
\end{longtblr}

\subsection{Model / Algorithm Development}
Providing properly configured infrastructure for building ML models 
ensures reproducibility and effective tracking. Unlike simple 
algorithms, ML models depend on multiple factors that influence their 
behavior, including the training data used, the computing environment, 
and choices made during development. To maintain clear references for 
different models, metadata is saved alongside each model, including 
the training datasets, environment details, source code, performance 
metrics, and training curves\footnote{Training curves show how a 
model's performance improves over the course of training, and are 
used to diagnose whether a model has been trained sufficiently.}.

\subsubsection{Model / Algorithm Development Steps}
Developing an ML model involves the following steps:
\begin{enumerate}
    \item \myuline{Select data and metric to optimize}: Reference 
    the datasets used for development. Choose a target variable to 
    predict and a measure of how well the model performs. Common 
    measures include accuracy, recall, F1 score, and cross-entropy, 
    chosen based on whether the prediction is continuous or 
    categorical.

    \item \myuline{Write tests}: Develop tests to verify the model 
    works correctly at each stage of development. These include unit 
    tests, integration tests, and end-to-end tests.

    \item \myuline{Create a baseline model}: Start with a simple 
    heuristic or non-ML calculation that serves as a benchmark 
    against which to compare future models.

    \item \myuline{Identify appropriate models to test}: Depending 
    on the problem, candidates may range from simpler approaches 
    such as boosted decision trees to more complex deep learning 
    architectures such as convolutional neural networks (CNNs) or 
    transformers. Other options include multi-layer perceptrons 
    (MLPs), variational autoencoders, and recurrent neural networks 
    (RNNs).

    \item \myuline{Train models and compare performance}: Compare 
    candidates to determine which one to proceed with. Key factors 
    include robustness, performance measures, training time, model 
    size, and complexity.

    \item \myuline{Fine-tune hyperparameters}: Tune the 
    configuration settings that govern how the model trains (as 
    distinct from the weights the model learns on its own). 
    Hyperparameters include learning rate, batch size, number of 
    training iterations, and regularization strength.

    \item \myuline{Tag model}: When the model is ready for further 
    development toward deployment, tag it as 
    ``request\_review''. This will alert administrators that a new 
    model is being developed, and they can advise the model builder 
    on the next steps for production.
\end{enumerate}

\subsubsection{Model / Algorithm Development Tools}
To effectively track experiments with various model architectures, it 
is important to record the code, package dependencies, training and 
test datasets, and the resulting performance using standard tracking 
and version control tools. While ad hoc approaches such as complex 
directory structures have been used in the past, adopting proper tools 
and best practices ensures that work is easily reproducible and 
shareable with all stakeholders.

Repositories and registries do not necessarily require separate tools. 
For example, a single platform like GitHub can serve multiple 
purposes if configuration and result files are kept within the same 
code repository. For more advanced needs, specialized solutions like 
MLFlow and the \textbf{Metadata Catalog} can be used to manage models 
and their associated metadata.

\begin{itemize}
    \item \textbf{Code Repository}, see Table~\ref{tab:tooling_code_repository}. This is where \myuline{tests} and \myuline{models} are version controlled.
    \item \textbf{Environment Registry}, see Table~\ref{tab:tooling_environment_registry}. This is possibly part of the \textbf{code repository}; its key function is to save the environment required to re-create the \myuline{models}.
    \item \textbf{Experiment Registry}, see Table~\ref{tab:tooling_experiment_registry}. This is used to document \myuline{models} and \myuline{hyberparameter} settings and compare them with each other. This will also be the tool used to \myuline{tag a model} for the next step of development.
\end{itemize}

\begin{longtblr}[
        caption={Tooling specification for \textbf{Code Repository}.}, 
        label={tab:tooling_code_repository}
    ]{
        hlines, vlines,
        cells = {valign=m,font=\footnotesize},
        column{1} = {bg=acornBlue, fg=white},
        column{1} = {0.85in,l,m},
        column{2} = {5.15in,l,m}
    }
\textbf{Core Functionality} & Keep code organized and version controlled. Enables automated testing [e.g. end-to-end, integration, unit], automatable deployments, collaborative development, and configurable user permissions.\\ 
\textbf{User Tools} & Users can create skeleton project with tests, environment, reference to datasets, deployment demo etc. Interfaces GUI and command line.\\
\textbf{Administrative Tools} & Define skeleton projects. Administrative scans such as to require test implementations, correct permissions, cyber security, etc.\\
\textbf{Integrations} & Store references to \textbf{Environment Registry} (for development and another for production), \textbf{Metadata Catalog} (for datasets used in creation of this model).
\end{longtblr}

\begin{longtblr}[
        caption={Tooling specification for \textbf{Environment Registry}.}, 
        label={tab:tooling_environment_registry}
    ]{
        hlines, vlines,
        cells = {valign=m,font=\footnotesize},
        column{1} = {bg=acornBlue, fg=white},
        column{1} = {0.85in,l,m},
        column{2} = {5.15in,l,m}
    }
\textbf{Core Functionality} & Keep track of model environments, ensure that the environments build. Store a defined list of all software packages and versions for a model as well as environment variables and secret handling.\\ 
\textbf{User Tools} & Export built environment using typical package managers [e.g. pip, venv, conda, uv]\\
\textbf{Administrative Tools} & Administrative scans for misuse of directory structures, package vulnerabilities. Define a package retention policy and automations.\\
\textbf{Integrations} & Accessible / callable from \textbf{Code Repository}. Interfaces to secure internal mirrors of requested packages.
\end{longtblr}

\begin{longtblr}[
        caption={Tooling specification for \textbf{Experiment Registry}.}, 
        label={tab:tooling_experiment_registry}
    ]{
        hlines, vlines,
        cells = {valign=m,font=\footnotesize},
        column{1} = {bg=acornBlue, fg=white},
        column{1} = {0.85in,l,m},
        column{2} = {5.15in,l,m}
    }
\textbf{Core Functionality} & Used for keeping track of model experiments during development. Experiment components: 1. tag to dataset in \textbf{Metadata Catalog}, 2. tag to code in \textbf{Code Repository}, 3. tag to environment in \textbf{Environment Registry}, 4. evaluation metrics (learning curves of train and validate datasets, performance on test sets), and 5. a unique identifier for the experiment. Ability to summarize and compare experiments \\ 
\textbf{User Tools} & GUI to browse experiments, compare experiments, share experiments, add comments, support tags defined by users and tags defined by administrators.\\
\textbf{Administrative Tools} & Administrator configurable status tags required for tracking [``trial'', ``best'', ``request\_review'', ``for\_deployment'']. Administrative scans for compliance (cyber security, etc.).\\
\textbf{Integrations} & Store references to \textbf{Code Repository} code tagged commits, \textbf{Environment Registry} environments, \textbf{Metadata Catalog} datasets. Automatable experiment creation and tagging.
\end{longtblr}

\subsubsection{Model / Algorithm Development Risk Mitigation}
These model and algorithm development tools mitigate the risks listed in Table~\ref{tab:mlops_model_development_system_risk_mitigations}.

\begin{longtblr}[ 
        caption={\textbf{Model / Algorithm Development} system risks and mitigations.}, 
        label={tab:mlops_model_development_system_risk_mitigations}
    ]{
        hlines, vlines,
        row{1} = {bg=acornBlue, fg=white}, 
        rowhead = 1,
        cells = {valign=m,font=\footnotesize},
        column{1} = {2.5in,l},
        column{2} = {3.5in,l}
    }
\textbf{Risk} & \textbf{Mitigation}\\*
Lack of reproducibility & The properly enforced use of the \textbf{Code Repository}, and \textbf{Environment Registry} will ensure that the models can be re-built and software dependencies can be iteratively upgraded. \\*
Lack of testing / benchmarking & Requiring automated testing in the \textbf{Code Repository} will ensure that the main code branch always works. The \textbf{Experiment Registry} will store benchmarking results along with references to the proper code and data.\\
Lack of oversight & Machine departments should be involved during this phase and definitely before entering the next phase. A sign-off approval process should be implemented.\\
Model accuracy degraded by edge optimization & Validation of the optimized model against the original on held-out test data is required before deployment, and results are version controlled as part of the edge model package.
\end{longtblr}

\subsection{Productionalize}\label{sec:mlops:productionalize}
Productionalization refers to the preparation, testing, and packaging 
of a model for deployment. While models are typically developed and 
trained on historical data, moving to production requires integrating 
with live data streams, processing data in real time, and running a 
streamlined version of the model optimized for generating predictions 
(known as inference). The model's outputs must also be connected to 
downstream systems to trigger the desired actions.

\subsubsection{Productionalization Steps}
To deploy an ML model onto a live system requires the following steps:
\begin{enumerate}
    \item \myuline{ETL Script}: A script that extracts live data 
    from the control system, transforms it (applying any algebraic 
    calculations the model requires), and loads it in the format 
    the model expects. ETL stands for ``extract, transform, load.'' 
    This script should be developed, tested, and version controlled.

    \item \myuline{Inference Model}: A streamlined version of the 
    model, optimized to generate predictions from live data. The 
    inference environment typically requires only a limited number 
    of software packages compared to the full development 
    environment.

    \item \myuline{Edge Model Preparation}: For models intended for field deployment, the inference model must optimized and validated for operation on the target edge hardware. This includes applying appropriate techniques such as quantization or architecture simplification, and verifying that model accuracy is preserved after optimization. The optimized model and its validation results should be version controlled alongside the standard inference model assets.

    \item \myuline{Delivery Scripts}: Scripts that send the model's 
    output to all required destinations. This may include 
    calculating and setting a process variable, sending it to a 
    monitor, and/or logging it.

    \item \myuline{Edge Delivery Configuration}: For field-deployed models, delivery includes configuring the model package for remote deployment to the field device, and defining the aggregate statistics and effectiveness measures that the field device will report upstream to the central control server for monitoring purposes.

    \item \myuline{Monitoring Scripts}: Scripts that track metrics 
    indicating whether the model continues to perform as expected. 
    For example, the distributions of input data and model 
    predictions should remain statistically consistent with the 
    training, validation, and test sets used to build the model.
\end{enumerate}

\subsubsection{Productionalization Tools}
The following tools support the productionalization process:
\begin{itemize}
    \item \textbf{Code Repository}, see Table~\ref{tab:tooling_code_repository}. This will be used to store, tag, and pull each of \myuline{ETL Script}, \myuline{Inference Model}, \myuline{Delivery Scripts}, and \myuline{Monitoring Scripts}.
    \item \textbf{Environment Registry}, see Table~\ref{tab:tooling_environment_registry}. This will be used to store, tag, and pull environments for \myuline{ETL Script}, \myuline{Inference Model}, \myuline{Delivery Scripts}, and \myuline{Monitoring Scripts}.
    \item \textbf{Model Package Registry}, see Table~\ref{tab:tooling_package_registry}. This is a more general version of the \textbf{experiment registry}. It includes deployment configurations and pipeline tooling for \myuline{ETL scripts} to prepare live data, hosting the \myuline{inference model} for generating predictions, connecting \myuline{monitoring scripts} with a \textbf{model monitoring, command, and control} application, and setting into motion resulting actions. It contains everything needed for deploying an integrated new functionality in the control system. For edge deployments, the optimized edge model and its validation results are also saved here.
\end{itemize}

\begin{longtblr}[
        caption={Tooling specification for \textbf{Model Package Registry}.}, 
        label={tab:tooling_package_registry}
    ]{
        hlines, vlines,
        cells = {valign=m,font=\footnotesize},
        column{1} = {bg=acornBlue, fg=white},
        column{1} = {0.85in,l,m},
        column{2} = {5.15in,l,m}
    }
\textbf{Core Functionality} & Database of model packages including \myuline{ETL Script}, \myuline{Inference Model}, \myuline{Delivery Scripts}, \myuline{Monitoring Scripts}, and their environments. Also supports tags defined by the owner or administrator-defined status tag [e.g. ``deployed'', ``retired'', ``dev'']\\ 
\textbf{User Tools} & GUI and/or command line tool to submit a package.\\
\textbf{Administrative Tools} & Ability to update tags, and produce audit summaries of the packages.\\
\textbf{Integrations} & Automatable tag updates, and deployment of packages.
\end{longtblr}

\subsubsection{Productionalization Risk Mitigation}
\begin{longtblr}[ 
        caption={\textbf{Productionalization} system risks and mitigations.}, 
        label={tab:productionalization_risk_mitigations}
    ]{
        hlines, vlines,
        row{1} = {bg=acornBlue, fg=white}, 
        rowhead = 1,
        cells = {valign=m,font=\footnotesize},
        column{1} = {2.5in,l},
        column{2} = {3.5in,l}
    }
\textbf{Risk} & \textbf{Mitigation}\\*
Lack of reproducibility and benchmarking. & \SetCell[r=3]{l}{The \textbf{Model Package Registry} and supporting tooling ensure that the model that will be deployed is buildable, and it keeps track of ingredients needed to reproduce the deployable version of the model, including: \myuline{ETL Script}, \myuline{Inference Model}, \myuline{Delivery Scripts}, \myuline{Monitoring Scripts}, and their environments}\\*
Accounting of components necessary for responsible deployment. & \\*
Grouping of components for containerized (Kubernetes) deployment and monitoring. & \\*
Lack of oversight & Machine departments should be involved before pursuing productionalization. A sign-off approval process should be implemented.
\end{longtblr}

\subsection{Operate}\label{sec:mlops:operate}
System operations includes everything needed to deploy, monitor, and 
control ML models within the accelerator control room. Introducing 
automation, whether powered by AI/ML or not, aims to enhance 
accelerator performance and the user experience for operators. 
Control room operators are critical front-line problem-solvers, and 
these automations are intended to provide greater context, insight, 
and support for decision-making. For example, the automations listed 
in Section~\ref{sec:AI_Prerequisites_and_Use-Cases} provide 
higher-fidelity alarms and more robust auto-tune scripts that perform 
reliably across a wider range of scenarios. 

Each automation should, 
at a minimum, provide clear and informative alarms for operators, 
along with interfaces for performance monitoring and command controls 
when necessary. Before productionalization and operation of any 
automation, well-defined expectations must be established in 
concurrence with machine experts and the operations department. 
Additionally, every automation must have designated primary and 
secondary owners who are responsible for ongoing maintenance and 
addressing any on-call issues.

The steps to deploy a model should be minimal for both centrally hosted models and for the central monitoring functions associated with edge-deployed models, for which a single step is needed.
For the field-deployed components of an edge automation, deployment procedures will depend on the target hardware and must be developed and documented as part of the productionalization process.
Because edge deployments require re-optimization and re-validation of the model before each redeployment, the retrain-to-redeploy cycle is longer than for centrally hosted models and should be accounted for in the maintenance plan for each automation.
\begin{enumerate}
    \item \myuline{Update model package tag to ``deployed''}: This 
    should automatically unpack and activate the model package 
    elements: \myuline{ETL Script}, \myuline{Inference Model}, 
    \myuline{Delivery Scripts}, and \myuline{Monitoring Scripts}.
\end{enumerate}

\subsubsection{Operation Tools}
The following tools are recommended to support operations of ML models:
\begin{itemize}
    \item \textbf{Model Monitoring, Command, and Control}, see Table~\ref{tab:tooling_mcc}. Once a \myuline{model package tag} is updated, this tool will be the main GUI for humans monitoring and controlling the state of the model.
\end{itemize}

\begin{longtblr}[
        caption={Tooling specification for \textbf{Model Monitoring, Command, and Control}.}, 
        label={tab:tooling_mcc}
    ]{
        hlines, vlines,
        cells = {valign=m,font=\footnotesize},
        column{1} = {bg=acornBlue, fg=white},
        column{1} = {0.85in,l,m},
        column{2} = {5.15in,l,m}
    }
\textbf{Core Functionality} & Display model monitoring metrics for operators and specialists. Display metadata about deployed model package. Enable actions such as: pause and resume (and as desired, an automated re-train action). Configurable autonomy levels for more advanced models.\\ 
\textbf{User Tools} & GUI in control system to perform the Core Functionality.\\
\textbf{Administrative Tools} & N/A\\
\textbf{Integrations} & Command logging with the \textbf{control system}.
\end{longtblr}

\subsubsection{Operation Risk Mitigation}
\begin{longtblr}[ 
        caption={\textbf{Operation} system risks and mitigations.}, 
        label={tab:operation_risk_mitigations}
    ]{
        hlines, vlines,
        row{1} = {bg=acornBlue, fg=white}, 
        rowhead = 1,
        cells = {valign=m,font=\footnotesize},
        column{1} = {2.5in,l},
        column{2} = {3.5in,l}
    }
\textbf{Risk} & \textbf{Mitigation}\\*
Lack of visibility into model performance. & \SetCell[r=2]{l}{The \textbf{Model Monitoring, Command, and Control} is a central and standardized hub for operators to observe and, if necessary, control ML models when they go into alarm. As models are improved and advance to higher levels of autonomy, the control and interaction with the ML model will change.}\\*
Difficult to use one-off expert systems.&\\*
Unreliable operations of ML models. & Models are documented and launched via this MLOps system.
\end{longtblr}

\subsection{MLOps Oversight Application}
Finally, an MLOps oversight application should be created that 
provides a window into the status of automations as they move through 
the MLOps pipeline. This application is a dashboard that receives 
push notifications from the tools described above. It would help the 
Controls Department shepherd automations through the process of 
ideation, data discovery, algorithm design, productionalization, and 
deployment. Optionally, the application could also receive push 
notifications from an approvals database to ensure the responsible 
machine and operations leaders are aware of new automations. The 
MLOps Oversight Application would provide observability and serve as 
a hub for governance of AI/ML development and deployment. For example, the concurrent deployment of interacting automations should be monitored and avoided, this could be a main function of the MLOps Oversight Application.

\section{Data Quality and Access}
\label{sec:Data_quality_and_access}
Fermilab's original accelerator systems were designed in a tightly 
integrated manner, similar to the movements of a complex wristwatch. 
Signals were passed directly between systems to orchestrate 
synchronous transfers of beam between machines. With the advent of computers 
and Fermilab's original CAMAC field bus system, this functionality 
expanded into the digital realm. However, the machines themselves 
remain fully resonant, depending on analog power systems provided by 
public utilities. Coordinating between these digital and analog 
domains presents ongoing opportunities for innovation and improvement.

ACORN can systematically redefine how data is collected and 
structured to better serve the needs of current and future machine 
experts. Bridging the analog and digital domains requires reliable, 
standardized timestamps generated as close as possible to the moment 
the data is acquired. Appropriate timestamps enable synchronization 
of events across different data sources, monitoring of signal and 
device integrity, and real-time processing for controls decisions. 
For example, precision time-stamping of RF and instrumentation 
systems reveals correlations that can be used to train ML models. 
Combining high-integrity timestamps with expanded signal bandwidth 
enables faster feedback algorithms across general accelerator 
systems. End users may care primarily about correlations between signals rather than the timestamps themselves, but high‑quality timestamps are essential to construct those correlations in the first place. A well-organized integration of both capabilities is needed 
to bridge the gap between the digital control system and the highly 
analog machines it controls.

\subsection{Users of Accelerator Data}
Accelerator complexes face a unique challenge in maintaining data 
integrity on timescales ranging from years to nanoseconds. By first 
identifying what systems and people use data from the accelerator, 
ACORN can define data quality prerequisites tailored to each 
stakeholder. This document focuses specifically on the stakeholders 
involved in designing, deploying, and monitoring AI/ML algorithms in 
the accelerator control system.

\begin{longtblr}[
        caption={Data stakeholders.}, 
        label={tab:data_stakeholders}
    ]{
        hlines, vlines,
        row{1} = {bg=acornBlue, fg=white}, 
        rowhead = 1,
        cells = {valign=m,font=\footnotesize},
        column{1} = {0.55in,c},
        column{2} = {0.9in,l},
        column{3} = {4.35in,l}
    }
\SetCell[c=2]{c}{\textbf{Stakeholder}} && {\textbf{Examples}}\\* 
\SetCell[r=3]{c} Systems
    & Hardware & Actuators: e.g. magnet power supplies, valves, pumps, klystrons.\newline Instrumentation: ex: resistive wall current monitors, beam position monitors.\\*
    & Firmware & Edge devices: e.g. ramp card settings for magnet power supplies, READS predictive loss system.\\*
    & Software & Edge device software. User Applications: parameter page, settings interfaces, synoptic displays, AI/ML diagnostics and controls.\\*  
\SetCell[r=3]{c} {Humans}
    & {Researching and Developing} & {Collecting data for offline analysis (e.g. correlating effects across the complex, building algorithms behind software and hardware solutions).}\\*
    & {Developing and Deploying} & {Transforming data pipelines for automations (e.g. implementing software and hardware solutions like feedback PID loops, regulation scripts, automated tune measurement, beam position monitors, AI/ML diagnostics and controls).}\\*
    & {Operating} & {Observing consolidated data for decision-making (e.g. explore data in ``fast-time'' plot, alarm page, synoptic displays).}
\end{longtblr}

\subsection{Data Quality}
Each stakeholder listed in Table~\ref{tab:data_stakeholders} may have 
different data quality requirements. For example, consider an operator 
in the Main Control Room monitoring the fluid level in a cooling tank, 
looking for trends over days or weeks to determine when to schedule a 
technician to refill it. If the data is collected every 10 minutes, 
occasional missing data points are unlikely to affect their decision. 
Similarly, minor timestamp jitter on the order of a minute would have 
little effect, but a jitter of 12 hours could lead the operator to 
refill earlier than necessary. A few inaccurate readings may be 
tolerable, but consistently low accuracy would prompt a more 
conservative response, such as refilling immediately. In general, the 
greater the uncertainty in the data, the more cautious the 
decision-making becomes.

This example highlights how data quality directly influences 
operational decisions. In ``Understanding data quality in a 
data-driven industry context: Insights from the 
fundamentals''~\cite{Fu2024}, the authors demonstrate a clear 
methodology to identify and define quality dimensions (accuracy, 
completeness, consistency, and timeliness) and specific metrics 
(mathematical calculations measuring each dimension). They emphasize 
that different dimensions and metrics will be relevant for different 
data uses. ACORN should apply these dimensions and metrics to 
determine quality standards for accelerator data. It should be noted 
that the timestamp associated with each data point is itself a data 
point and needs appropriate quality dimensions and metrics as well.

As the adage goes, ``garbage in, garbage out.'' No matter the 
sophistication of the AI/ML algorithm, its effectiveness fundamentally 
depends on the quality of the data it is trained on.

\subsection{Data Prerequisites}
The following prerequisites are suggested to improve overall data quality and trust.

\subsubsection{Ontology of a Datapoint}
Each data point recorded in the control system should include the 
following features to make it easy to understand and use.

\begin{enumerate}
    \item \myuline{Measurement is defined}: How are signals 
    calibrated into physical units? What is the measurement 
    uncertainty? Currently, in Fermilab's control system, physical 
    units are called ``Engineering Units,'' and digitized signals 
    present as raw integer values, ``Raw''. For most control applications, it 
    is useful to refer to physical units such as volts, proton 
    intensity, or millimeters.

    \item \myuline{Time of the measurement is defined}: How is the 
    timestamp calculated? Does it represent the start, end, or 
    middle of the measurement? What is the uncertainty of the time? 
    At present, Fermilab utilizes no fewer than five different 
    time-stamping methods. To simplify calculations, it is 
    advantageous to adopt a single Grandmaster time source.

    \item \myuline{Context is defined}: What is the state of the 
    accelerator at the time the measurement is made? Where in the 
    timeline sequence does it fall? What came before and after? What 
    were the accelerator settings and environmental conditions? 
    Minor changes can drastically impact downstream readings, so it 
    is important to capture accelerator run parameters in order to 
    distinguish between intentional and ambient changes in a data 
    point.

    \item \myuline{Payload is defined}: What is the shape, format, 
    and type of the data? How should it be unpacked? Visibility into 
    these parameters allows developers to readily use data without 
    relying on documentation or ``tribal knowledge.''
\end{enumerate}

\subsubsection{Data Quality Enforcement}
Data integrity and trustworthiness can be achieved through quality 
monitoring, metadata reporting, and parameter hashing.
\begin{enumerate}
    \item \myuline{Data quality monitoring}: Developing and 
    maintaining a framework for marking periods of data as ``good'' 
    when measurements, timestamps, context, and payload are within 
    tolerance. This allows developers to readily identify which data 
    is suitable for use in model training and evaluation.

    \item \myuline{Metadata reporting}: Many users perform the same 
    calculations on raw data independently. By computing summary 
    statistics and reporting data context in advance, redundant 
    processing is reduced and calculations are standardized across 
    applications.

    \item \myuline{Parameter hashing}: Generating a compact 
    fingerprint (called a hash) for key parameters such as 
    calibration coefficients and run mode. This allows users to 
    quickly assess whether two data points were taken under the same 
    conditions, or whether external changes have affected the 
    readings, without tracking every settings change through time.
\end{enumerate}

\subsubsection{Data Accessibility}\label{sec:data_access}
Data accessibility focuses on timeliness and ease of use. Each user 
of the data should have access to well-documented and timely data. 
System users (hardware, firmware, and software) will have different 
access needs than human users (researchers, developers, and 
operators). The 
tools needed to support data access for AI/ML researchers are described next.

Data is handled by two systems at Fermilab, each handle different types of accelerator data. The first is the new control system that 
replaces ACNET. ACNET's primary purpose was the transmission of 
``live signals'' for real-time control and monitoring. The 
replacement must not only fulfill this role but also function as a 
reliable data acquisition system, capable of collecting and 
timestamping data to support AI/ML automation development.
The second system is being developed and will be deployed through a 
series of independent Accelerator Improvement Projects (AIPs).

Subsets of the proposed AIPs are targeted at improving beam 
instrumentation. The first is already underway, updating beam 
position monitors (BPMs) in the 8-GeV Transfer Line to prepare for 
high-intensity PIP-II and LBNF operations. The new instruments will 
use a field-based, high-frequency data framework built on Redis. 
Redis serves as device-side infrastructure that supports flexible, 
software-based transformations of dense, raw instrument data. For 
example, raw digitized signals from BPMs are fed to edge-compute 
racks where they are transformed into meaningful quantities such as 
beam position. This processed data is then relayed by the control 
system for central monitoring, decision-making, and archiving. 
Bridging the gap between AI/ML researchers and this high-frequency 
instrument data is in ACORN's interest, as it can provide a uniform 
experience for accessing and archiving this data.

The following tool specifications describe how data is collected 
through the future control system 
(\mbox{Table~\ref{tab:tooling_control_system_data_interface}}) and 
through the Redis-based device-side framework 
(Table~\ref{tab:tooling_instrumentation_data_interface}).

\begin{longtblr}[
        caption={Tooling specification for \textbf{Control System Data Interface}.}, 
        label={tab:tooling_control_system_data_interface}
    ]{
        hlines, vlines,
        cells = {valign=m,font=\footnotesize},
        column{1} = {bg=acornBlue, fg=white}, 
        column{1} = {0.85in,l,m},
        column{2} = {5.15in,l,m}
    }
\textbf{Core Functionality} & Standardized methodology to select live and historic data following practices in Section~\ref{sec:Data_quality_and_access}. \\ 
\textbf{User Tools} & Interface to request data from the control system and standard instrumentation. Provide clear definitions of all data fields.\\
\textbf{Administrative Tools} & Ability to throttle or require approval when system limitations are reached.\\
\textbf{Integrations} & Reads from the \textbf{control system}. Writes to \textbf{Dataset Storage Volume}.
\end{longtblr}

\begin{longtblr}[
        caption={Tooling specification for \textbf{Instrumentation Data Interface}.}, 
        label={tab:tooling_instrumentation_data_interface}
    ]{
        hlines, vlines,
        cells = {valign=m,font=\footnotesize},
        column{1} = {bg=acornBlue, fg=white},
        column{1} = {0.85in,l,m},
        column{2} = {5.15in,l,m}
    }
\textbf{Core Functionality} & Standardized methodology to select streaming data from instruments. \\ 
\textbf{User Tools} & Interface to request instrumentation data from various instruments (BPM waveforms, RWM waveforms, etc.), and user-defined in-situ transformations. Clear definitions of all data fields.\\
\textbf{Administrative Tools} & Ability to throttle or require approval if system limitations are met. Permission controls to suggest consulting the instrumentation engineers for advisement for non-standard requests.\\
\textbf{Integrations} & Reads from the \textbf{Instrumentation's Redis Cache}. Writes to \textbf{Dataset Storage Volume}.
\end{longtblr}

Currently, only system experts and developers know the details of 
accessing data from each source, and they rely on this knowledge when 
developing software. Researchers and operators need consolidated 
pathways for collecting data that reduce the specialized knowledge 
previously required. This is commonly referred to as ``data 
democratization''.

AI/ML researchers need to be able to save selected data from three 
sources: 1.~Redis, 2.~ACORN's Data Cache, and 3.~the Dataset Storage 
Volume, which may reside inside or outside the controls network firewall 
(Table~\ref{tab:tooling_dataset_storage_volume}). ACORN's Data Cache 
will provide access to pre-configured data streams constructed by the 
front-ends, formatted in ``engineering units.'' Access to raw readout 
streams directly from Redis will need to be configured and monitored 
with the help of Instrumentation Department engineers to ensure the 
data meets the researcher's needs. This supervision is needed because 
Redis data can require extremely high throughput and is not practical 
to collect and store indefinitely.

Access to Redis data streams is a critical capability for AI/ML 
development and prototyping. It allows researchers to collect data 
that was previously difficult to obtain, as it required dedicated 
instrumentation engineering time. With this data, ideas can be 
prototyped before committing to major engineering effort.

Finally, the Dataset Storage Volume should be accessible from 
wherever researchers are doing analysis, building models, and sharing 
data with collaborators. Because it is used for offline study, it 
should be accessible outside the controls network firewall and be easily 
accessed from convenient compute environments such as Fermilab's 
Elastic Analysis Facility.

\paragraph{\textbf{Usability}} An important enhancement opportunity 
for data access is improved usability, particularly around 
timestamps. Nearly all AI/ML users must perform the same 
computationally intensive step: aligning sparse time-series data from 
many devices into a single time-aligned table. The resulting table, 
used for analysis and model building, is typically stored as a CSV or 
dataframe in which each column is a measurement from a different 
device and each row represents a moment in time. \textbf{The 
practicality of supplying pre-aligned time-series data natively 
should be studied.}

\begin{enumerate}
    \item \myuline{Easy to use}: Data must be straightforward to 
    retrieve, understand, and merge. For real-time data, this can be 
    accomplished through shared memory infrastructure such as 
    Instrumentation's Redis cluster or ACORN's Data Cache. For 
    historical data, ACORN's Data Store should provide accessible, 
    well-documented archives.

    \item \myuline{Easy to share}: To support collaboration and 
    reduce rework, a centralized repository is needed so that others 
    can find, use, and understand prior work and datasets. This 
    repository, the Dataset Storage Volume, should be accessible to 
    collaborators such as experimenters and other accelerator 
    facilities outside the controls firewall.
\end{enumerate}

\subsection{Data Quality Risk Mitigation}
Implementing strong data quality practices helps mitigate several 
risks, including wasted time and resources, allowing researchers to 
focus on operational improvements and enabling faster onboarding of 
collaborators. Table~\ref{tab:dataset_quality_risk_mitigations} 
illustrates the consequences of poor data quality across key 
dimensions. 

\begin{longtblr}[ 
        caption={Data quality risks and consequences.}, 
        label={tab:dataset_quality_risk_mitigations}
    ]{
        hlines, vlines,
        row{1} = {bg=acornBlue, fg=white}, 
        rowhead = 1,
        cells = {valign=m,font=\footnotesize},
        column{1} = {1.4in,l},
        column{2} = {4.6in,l}
    }
\textbf{Risk} & \textbf{Consequences}\\
Poor time & Timestamps for different measurements are not accurate or well understood, making it very difficult to join data streams for correlation analyses.\\
Poor measurements & If basic expectations about the reliability of reported measurements are not available, every analysis would have to begin by assessing this independently.\\
Poor context & Without knowledge of the machine state when measurements are taken, researchers and developers will not be able to easily explain data drifts.\\
Poor payload description and bookkeeping & Every researcher would have to determine the data format independently by tracking down the engineers who configured the communication signals.\\
Poor quality monitoring & Researchers and developers would have to devise their own quality metrics before building a model.\\
Poor metadata reporting & Researchers and developers would not be able to reuse datasets because they would not know the circumstances or devices included in each dataset.\\
Poor accessibility & Researchers and developers would not be able to perform analyses efficiently.
\end{longtblr}

\clearpage 
\section{AI Prerequisites and Use Cases}
\label{sec:AI_Prerequisites_and_Use-Cases}
\begin{figure}[bhtp!]
    \centering
    \Large{\textbf{Fermilab Accelerator Complex Automation Taxonomy}}
    \includegraphics[width=1\linewidth]{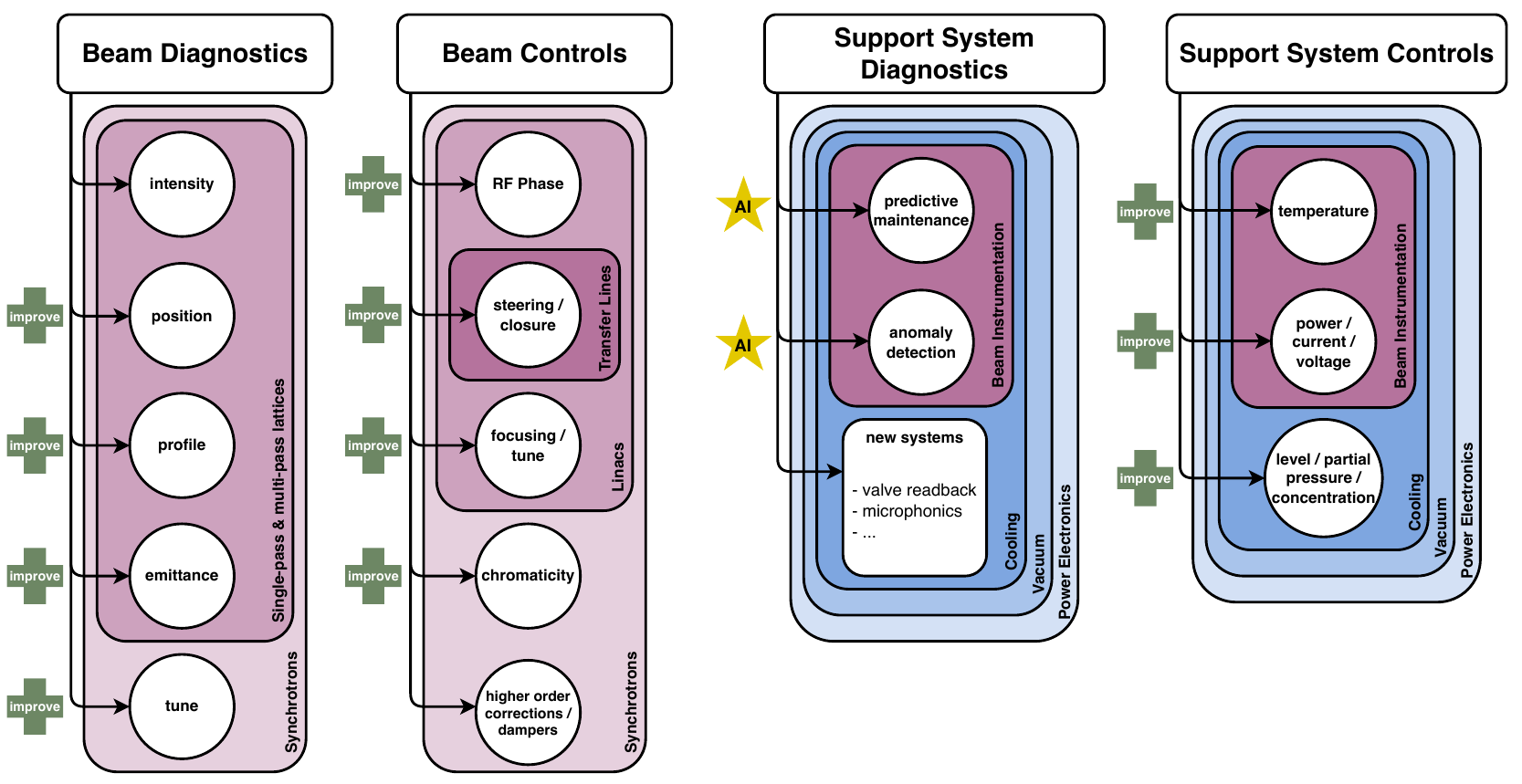}
    \caption{Taxonomy of automation use cases for diagnostics and controls for both beam and support systems. Shaded boxes represent the parts of the accelerator complex that use these diagnostics and controls. Stars and pluses indicate where AI/ML is necessary or can lead to operational improvements.}
    \label{fig:Automation_taxonomy}
\end{figure}

Figure~\ref{fig:Automation_taxonomy} shows a taxonomic summary of automation 
use cases throughout Fermilab's accelerator complex. 
Automations are listed in four columns, the first two refer to beam diagnostics and controls, and the second towo refer to diagnostics and controls for support systems. AI/ML will 
initially have the greatest impact on Support System Diagnostics, 
including predictive maintenance and anomaly detection (marked with 
stars in Figure~\ref{fig:Automation_taxonomy}). All other use cases 
(marked with pluses in Figure~\ref{fig:Automation_taxonomy}) will improve as users explore accessible accelerator data (see Section~\ref{sec:Data_quality_and_access}), better explore, develop, deploy and track their automations (see Section~\ref{sec:Lifecycle_Management_of_Automations}). Some automations may be sufficient as-is and may not be improvable with AI/ML or its infrastructure, such as intensity diagnostics and dampers.

\subsection{Prerequisites to Enable AI/ML Automations}
Table~\ref{tab:automation_prerequisites} lists the prerequisites 
needed to enable AI/ML-based automations and decision-making. The details of Prerequisites for Operations are calculable from the use cases in the next subsection. The pre-requisites for development are covered by the MLOps infrastructure, Section~\ref{sec:Lifecycle_Management_of_Automations}, and 
high-quality data, Section~\ref{sec:Data_quality_and_access}.

\begin{longtblr}[
        caption={Prerequisites to enable AI/ML-based decision making and automations.}, 
        label={tab:automation_prerequisites}
    ]{
        hlines, vlines,
        row{1} = {bg=acornBlue, fg=white, font=\bfseries}, 
        rowhead = 1,
        column{1} = {0.4in,c,m},
        column{2} = {5.5in,l,m}
    }
\SetCell[c=2]{l}{Requirement ID \& Description}&\\* 
\SetCell[c=2]{l}{\textbf{R.1. Prerequisites for Operations} -- The deployed model has:}\\*
R.1.1. & Access to incoming data, readbacks, statuses, settings, and control\\*
R.1.2. & Access to model package, including thresholds for metrics\\*
R.1.3. & Adequate CPU, memory, GPU, and networking to deliver timely actions\\*
R.1.4. & Model performance monitoring visuals and tracking database\\*
R.1.5. & Connection to systems for alarms, recommendations, and other automated actions\\
\SetCell[c=2]{l}{\textbf{R.2. Prerequisites for Development} -- Development engineers have:}\\*
R.2.1. & Access to historical (user-friendly) incoming data, readbacks, statuses, settings, and control\\*
R.2.2. & Access to research compute cluster with adequate CPUs, GPUs, and software\\*
R.2.3. & Easy method to access historical data from the research compute\\*
R.2.4. & Flexible software ecosystem that supports algorithm development and tracking tests\\*
R.2.5. & Documented productionalization method (how to prepare models for deployment)\\*
R.2.6. & Ability to request temporary logging of atypical data (waveforms, intermediate values from field equipment, readout at a custom trigger/delay, etc.)
\end{longtblr}

\subsection{Example Use Cases} \label{sec:AI_Use-Cases:Examples}
Listed in this section's tables are use cases collected from multiple 
rounds of interviews with departments across Fermilab's Accelerator 
Directorate. Implementation of these automations is not in scope for 
ACORN; however, the Controls Department AI/ML Group will assist 
system experts in developing them. ACORN aims to deliver a control 
system that provides experts with the data and tools they need to 
research and implement these automations. Before implementation, 
expectations should be agreed upon by system experts and the 
operations department, and each automation must have an owner who is 
responsible for maintenance and on-call issues. These automations 
give a sense of the technical requirements for ACORN's AI-ready 
control system.

\begin{figure}[ht]
    \centering
    \includegraphics[width=1\linewidth]{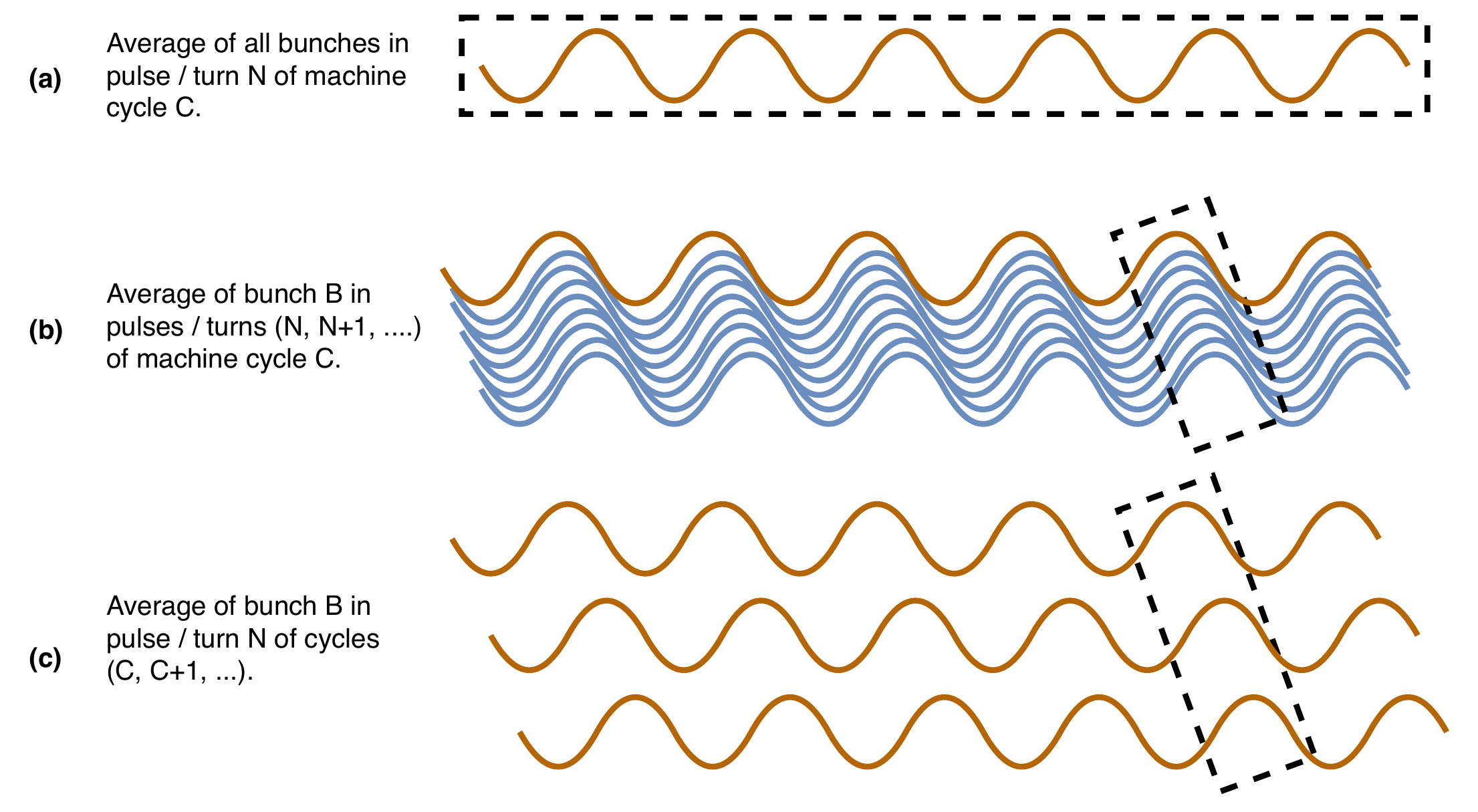}
    \caption{To get sufficient statistics for a measurement, beam instrumentation devices need to integrate signals from multiple bunches. (a) is the manner in which current measurements are usually made, (b) and (c) show alternative integrations which can be useful for understanding the accelerator.}
    \label{fig:Signals}
\end{figure}

To make assumptions clearer, resolutions in the use case tables are 
provided as relative measurements; for example, ``super cycle'' 
instead of 1~second and ``Booster beam cycle'' instead of 66.6 or 
50~microseconds. Many signals are integrated over an entire pulse (or 
turn) to accumulate adequate statistics for a measurement. A new 
request from users is to allow flexibility in how signals are 
integrated. For example, the first or last bunch of a pulse (or turn) 
may be more lossy, and users want the ability to resolve the 
characteristics of individual bunches averaged over many pulses or 
turns, see Figure~\ref{fig:Signals}.

\begin{figure}
    \centering
    \includegraphics[width=\linewidth]{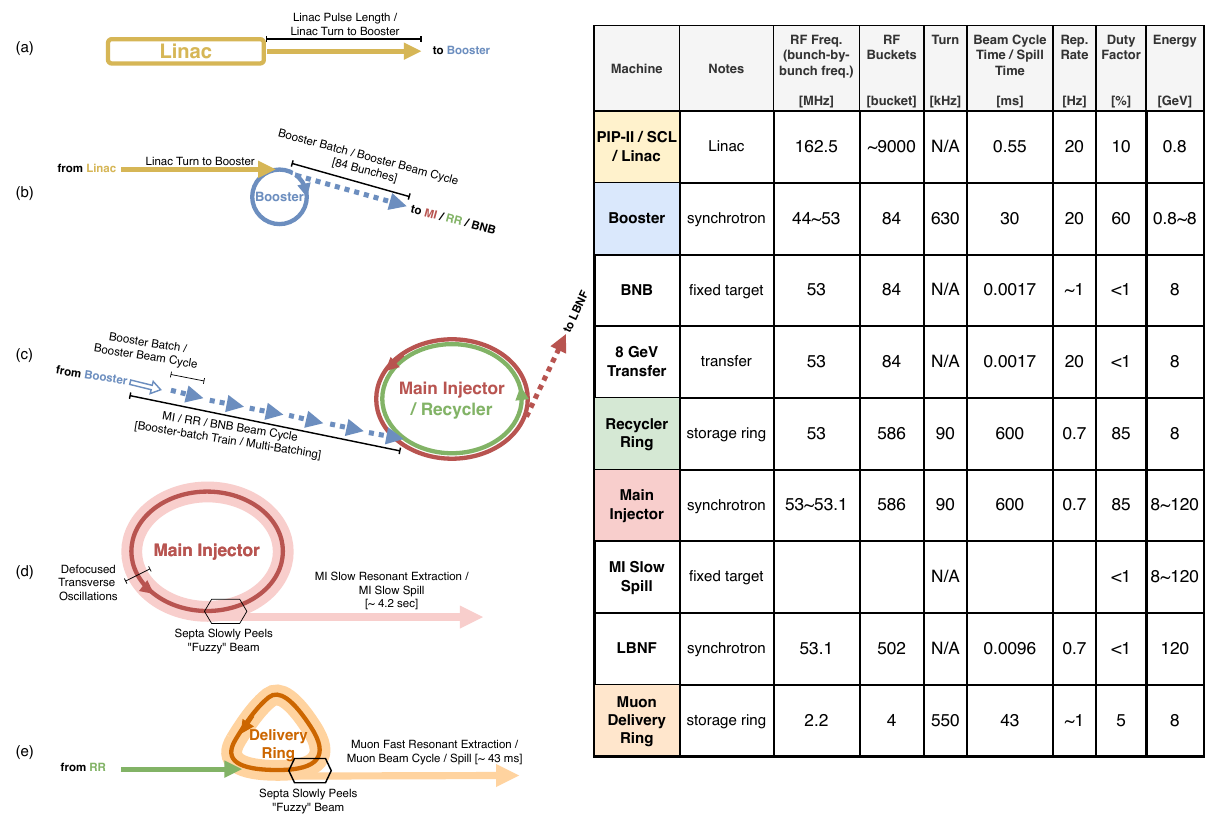}
    \caption{Right, shows table of parameters for the machines. Left, shows some facility-specific nomenclature: (a) Linac pulse length is the time-of-travel from the head to tail of the Linac pulse to Booster. (b) Currently, the Booster delivers a batch (up to 84 bunches) to MI / RR / BNB. (c) Multi-batch injection into MI / RR, also known as a Booster-batch train. (d) MI slow resonant extraction to Switchyard, also known as a slow-spill. (e) Fast resonant extraction of protons from the Muon Delivery Ring to the Mu2e target, solenoid, and detector.}
    \label{fig:nomenclature}
\end{figure}

The following use cases are organized into sections for beam and 
support systems, and each of those are further organized into diagnostic or control automations. 
Figure~\ref{fig:nomenclature} provides a visual guide to beam 
terminology found in the use cases. It should be noted that many of 
the automations require new instrumentation or controls that are 
outside ACORN's scope. The aspects within ACORN's scope include 
infrastructure (bandwidth, storage, compute) and software (user 
applications and the MLOps software stack).

\subsubsection{Automations in Beam Diagnostics}
Beam diagnostics automations cover the hardware and software used to compute quantities that describe particle beams. First-order quantities 
describe the beam centroid: (a)~momenta $p_x$, $p_y$, $p_z$; 
(b)~position $z(t)$, $x$, $y$; and (c)~angles $x'$, $y'$. 
Second-order quantities describe the distribution of particles: 
(a)~momentum spread; (b)~bunch length; (c)~beam size; (d)~beam 
divergences; and other correlations. Some quantities require the beam 
to be intentionally perturbed and the resulting measurements used in 
calculations, such as measuring the tune in circular machines.
\begin{itemize}
    \item Table~\ref{tab:beam_characterization}: Characterization (centroid and distribution in coordinate and momentum space)
    \item Table~\ref{tab:beam_anomaly}: State-based anomaly monitoring
    \item Table~\ref{tab:beam_tune}: Tune measurement
\end{itemize}

\begin{longtblr}[ 
        caption={Beam Diagnostics: Characterization (distribution in coordinate and momentum space).}, 
        label={tab:beam_characterization}
    ]{
        hlines, vlines,
        row{1} = {bg=acornBlue, fg=white, font=\bfseries}, 
        rowhead = 1,
        cells = {valign=m,font=\footnotesize},
        column{1} = {2in,l},
        column{2} = {1.6in,l},
        column{3} = {2.2in,l}
    }
\textbf{Machines} & \textbf{Acquisition Resolution} & \textbf{Description}\\*
Linac/PIP-II & \SetCell[r=7]{l} bunch-by-bunch\footnotemark[1] per pulse / machine's cycle time\footnotemark[1]& \SetCell[r=7]{l} Monitor reconstructed phase space using beam instrumentation, simulation, and lattice to ensure high beam transmission throughout the accelerator. Centrally deployed.\\*
Booster & & \\*
Recycler Ring & & \\*
Main Injector & & \\*
Muon Delivery Ring & & \\*
All transfer lines & & \\*
External beamlines (including LBNF) & & 
\end{longtblr}
\footnotetext[1]{See Figure~\ref{fig:nomenclature} where each machine's turn frequency, RF frequency (bunch-by-bunch), and beam cycle/spill time is listed.}

\begin{longtblr}[ 
        caption={Beam Diagnostics: State-based anomaly monitoring.}, 
        label={tab:beam_anomaly}
    ]{
        hlines, vlines,
        row{1} = {bg=acornBlue, fg=white, font=\bfseries}, 
        rowhead = 1,
        cells = {valign=m,font=\footnotesize},
        column{1} = {2in,l},
        column{2} = {1.6in,l},
        column{3} = {2.2in,l}
    }
\textbf{Machines} & \textbf{Acquisition Resolution} & \textbf{Description}\\*
Linac/PIP-II & \SetCell[r=7]{l} before machine's next beam cycle\footnotemark[1] (fast), 5-minute+ trend (slow) & \SetCell[r=7]{l} Statistical monitoring  of phase space, timing, and intensity (as a complement to the Machine Protection Systems and Beam Permit Systems). Use a state machine to define appropriate time-dependent alarm properties. Centrally deployed.\\
Booster & &  \\*
Recycler Ring & &  \\*
Main Injector & &  \\*
Muon Delivery Ring & &  \\*
All transfer lines & &  \\*
External beamlines (including LBNF) & &  
\end{longtblr}

\begin{longtblr}[ 
        caption={Beam Diagnostics: Tune measurement.}, 
        label={tab:beam_tune}
    ]{
        hlines, vlines,
        row{1} = {bg=acornBlue, fg=white, font=\bfseries}, 
        rowhead = 1,
        cells = {valign=m,font=\footnotesize},
        column{1} = {2in,l},
        column{2} = {1.6in,l},
        column{3} = {2.2in,l}
    }
\textbf{Machines}  & \textbf{Acquisition Resolution} & \textbf{Description}\\*
Booster & \SetCell[r=4]{l} upon request ($\approx$ 4 hours) per machine's beam cycle\footnotemark[1] & \SetCell[r=4]{l} Incorporate an automated procedure to ping a beam, and observe changes in beam position throughout a synchrotron to determine the beam's tune. Centrally deployed.\\*
Recycler Ring & &  \\*
Main Injector & &  \\*
Muon Delivery Ring & &  
\end{longtblr}
\footnotetext[1]{See Figure~\ref{fig:nomenclature} where each machine's turn frequency, RF frequency (bunch-by-bunch), and beam cycle/spill time is listed.}

\subsubsection{Automations in Beam Control}
Beam control use cases involve the hardware and software needed to 
adjust and regulate beam optics and timing. Although beam behavior 
can be sensitive to environmental and support system factors (such as 
vacuum quality), this section focuses on automations whose primary 
function is to adjust the beam lattice and timing.
\begin{itemize}
    \item Table~\ref{tab:beam_rf_manipulation}: RF manipulations
    \item Table~\ref{tab:beam_special_transfers}: Special transfer automations
    \item Table~\ref{tab:beam_steering_closure}: Steering and closure
    \item Table~\ref{tab:beam_focus_tune}: Focus and tune control
    \item Table~\ref{tab:beam_chromaticity_control}: Higher order corrections -- Chromaticity
    \item Table~\ref{tab:beam_dampening}: Higher order corrections -- Dampening of resonances and instabilities
    \item Table~\ref{tab:beam_collimation}: Collimation
    \item Table~\ref{tab:beam_general_control}: General beam automations
\end{itemize}

\begin{longtblr}[ 
        caption={Beam Control: RF manipulation.}, 
        label={tab:beam_rf_manipulation}
    ]{
        hlines, vlines,
        row{1} = {bg=acornBlue, fg=white, font=\bfseries}, 
        rowhead = 1,
        cells = {valign=m,font=\footnotesize},
        column{1} = {0.9in,l},
        column{2} = {2.15in,l},
        column{3} = {2.75in,l}
    }
\textbf{Beam Goal} & \textbf{Acquisition Resolution} & \textbf{Description}\\*
Linac/PIP-II RF Efficiency Regulation & piezoelectric tuner response time (fast), water temperature response time (slow) as needed when RF transmission dips & Ensure the energy supplied to the RF cavity is applied to acceleration, and not dissipated. Centrally deployed.\\

Linac/PIP-II RF Matching & bunch-by-bunch\footnotemark[1] upon user request (slow, as needed when beam profile changes) & Ensure the Twiss parameters match at RF cavities and BPMs during PIP-II operation (commissioning and recovery). Save and use this data to refine lattice model. Centrally deployed.\\

Booster Phase Locking for MI/RR & Booster's bunch-by-bunch\footnotemark[1] per <1~ms cogging operation & Ensure that the RF phase of Booster matches that of the downstream MI/RR so as to not spill protons. Centrally deployed.\\

Booster Cogging for MI/RR & MI/RR RF\footnotemark[1] injection-by-injection per Booster beam cycle\footnotemark[1] & Ensure that subsequent Booster extractions are precisely timed into a specific RF bucket for MI/RR. Centrally deployed.\\

Bunch Rotation in Booster & up to 4~GHz raw data upon user request (based on beam loss and periodically $\approx$4 hours) for a 30~ms Booster acceleration cycle & Quick scan of bunch rotation to select settings for good slip-stacking. Centrally deployed.
\end{longtblr}\footnotetext[1]{See Figure~\ref{fig:nomenclature} where each machine's turn frequency, RF frequency (bunch-by-bunch), and beam cycle/spill time is listed.}

\begin{longtblr}[ 
        caption={Beam Control: Special transfers.}, 
        label={tab:beam_special_transfers}
    ]{
        hlines, vlines,
        row{1} = {bg=acornBlue, fg=white, font=\bfseries}, 
        rowhead = 1,
        cells = {valign=m,font=\footnotesize},
        column{1} = {1.8in,l},
        column{2} = {1.3in,l},
        column{3} = {2.7in,l}
    }
\textbf{Beam Goal} & \textbf{Acquisition Resolution} & \textbf{Description}\\*
Septa Alignment for MI & \SetCell[r=3]{l} user defined ($\approx$4 hours) & \SetCell[r=2]{l} Correctly align relative postion of septa and beam for efficient split and correct beam intensities. Centrally deployed.\\*

Septa Alignment for Fixed Target &  & \\*

Resonant Extraction: MI to SY & MI/RR turn-by-turn\footnotemark[1] per SY spill\footnotemark[1] & Fine-control of power supplies would allow quad changes during slow-spill cycle. Centrally deployed.\\*

Resonant Extraction: Muon Delivery Ring to Mu2e &  Muon Delivery Ring's bunch-by-bunch\footnotemark[1] per Muon Campus spill\footnotemark[1] & Regulate extraction during Muon Campus beam cycles to ensure smooth fast-resonant extraction. \textbf{Deployed in field}, at machine; configurable centrally.
\end{longtblr}

\begin{longtblr}[ 
        caption={Beam Control: Steering and closure.}, 
        label={tab:beam_steering_closure}
    ]{
        hlines, vlines,
        row{1} = {bg=acornBlue, fg=white, font=\bfseries}, 
        rowhead = 1,
        cells = {valign=m,font=\footnotesize},
        column{1} = {1.95in,l},
        column{2} = {1.3in,l},
        column{3} = {2.55in,l}
    }
\textbf{Machine} & \textbf{Acquisition Resolution} & \textbf{Description}\\*
Linac/PIP-II & \SetCell[r=3]{l} bunch-by-bunch\footnotemark[1] averaged per pulse / machine's beam cycle\footnotemark[1] & \SetCell[r=3]{l}\textbf{Steering:} Adjust dipoles and trims to keep beam in ideal orbit / on target. Centrally deployed.\\
All transfer lines & & \\
External beamlines (including LBNF) & & \\
Booster \& ORBUMP/H$^{-}$ Stripping & \SetCell[r=4]{l} bunch-by-bunch\footnotemark[1] and turn-by-turn\footnotemark[1] averaged per machine's beam cycle\footnotemark[1] &\SetCell[r=4]{l}{\textbf{Injection Closure:} Fine-tune adjustments to place beam into ideal orbit. Centrally deployed. \newline\newline \textbf{Steering:} Adjust dipoles and trims to keep beam in ideal orbit / on target. Centrally deployed.}\\
Recycler Ring & & \\
Main Injector & & \\
Muon Delivery Ring & & 
\end{longtblr}

\begin{longtblr}[ 
        caption={Beam Control: Beam focusing and tune control.}, 
        label={tab:beam_focus_tune}
    ]{
        hlines, vlines,
        row{1} = {bg=acornBlue, fg=white, font=\bfseries}, 
        rowhead = 1,
        cells = {valign=m,font=\footnotesize},
        column{1} = {2in,l},
        column{2} = {1.6in,l},
        column{3} = {2.2in,l}
    }
\textbf{Machine} & \textbf{Acquisition Resolution} & \textbf{Description}\\*
Linac/PIP-II &\SetCell[r=3]{l} bunch-by-bunch\footnotemark[1] averaged per pulse / machine's beam cycle\footnotemark[1] & \SetCell[r=3]{l}\textbf{Focusing in single-pass machines:} Adjust quadrupoles and trims to appropriately control transverse beam spread. Centrally deployed.\\*
All transfer lines & & \\*
External beamlines (including LBNF) & & \\
Booster & \SetCell[r=4]{l} bunch-by-bunch\footnotemark[1] and turn-by-turn averaged per machine's beam cycle\footnotemark[1] &\SetCell[r=4]{l} \textbf{Tune adjustment in synchrotrons:} Adjust quadrupoles and trims to appropriately control transverse beam spread. Centrally deployed.\\*
Recycler & &  \\*
Main Injector & &  \\*
Muon Delivery Ring & &  
\end{longtblr}
\footnotetext[1]{See Figure~\ref{fig:nomenclature} where each machine's turn frequency, RF frequency (bunch-by-bunch), and beam cycle/spill time is listed.}

\begin{longtblr}[ 
        caption={Beam Control: Higher order corrections -- Chromaticity.}, 
        label={tab:beam_chromaticity_control}
    ]{
        hlines, vlines,
        row{1} = {bg=acornBlue, fg=white, font=\bfseries}, 
        rowhead = 1,
        cells = {valign=m,font=\footnotesize},
        column{1} = {1.1in,l},
        column{2} = {1.6in,l},
        column{3} = {3.1in,l}
    }
\textbf{Machine} & \textbf{Acquisition Resolution} & \textbf{Description}\\*
Booster & \SetCell[r=4]{l} Machine beam cycle-by-cycle\footnotemark[1] upon user request and periodically ($\approx$ 4 hours) & \SetCell[r=4]{l} Adjust sextupoles to automate chromaticity control, mitigate instabilities without excessive chromaticity. Expert automations based on BPMs, and allowable operator windows based on BLMs. Centrally deployed.\\*
Recycler Ring & & \\*
Main Injector & & \\*
Muon Delivery Ring & &
\end{longtblr}

\begin{longtblr}[ 
        caption={Beam Control: Higher order corrections -- Dampening of resonances and instabilities.}, 
        label={tab:beam_dampening}
    ]{
        hlines, vlines,
        row{1} = {bg=acornBlue, fg=white, font=\bfseries}, 
        rowhead = 1,
        cells = {valign=m,font=\footnotesize},
        column{1} = {1.1in,l},
        column{2} = {2.1in,l},
        column{3} = {2.6in,l}
    }
\textbf{Machine} & \textbf{Acquisition Resolution} & \textbf{Description}\\*
Booster & \SetCell[r=4]{l}bunch-by-bunch\footnotemark[1] per machine's beam cycle\footnotemark[1] upon user request and periodically ($\approx$ 4 hours) & \SetCell[r=4]{l} Maintain proper configuration of damper system, and create flexibility for easier modifications. \textbf{Deployed in field}, at machine; configurable centrally.\\*
RR & & \\*
MI & & \\*
Muon Delivery Ring & & 
\end{longtblr}

\begin{longtblr}[ 
        caption={Beam Control: Collimation.}, 
        label={tab:beam_collimation}
    ]{
        hlines, vlines,
        row{1} = {bg=acornBlue, fg=white, font=\bfseries}, 
        rowhead = 1,
        cells = {valign=m,font=\footnotesize},
        column{1} = {2in,l},
        column{2} = {1.8in,l},
        column{3} = {2in,l}
    }
\textbf{Machine} & \textbf{Acquisition Resolution} & \textbf{Description}\\*
Linac/PIP-II & \SetCell[r=7]{l} beam cycle-by-cycle (or pulse-by-pulse, or spill-by-spill)\footnotemark[1] upon user request ($\approx$ every 4 hours) & \SetCell[r=7]{l} Based on instrumentation, modify beam collimation by either moving collimators (slow), or steering beam into collimators (fast). Centrally deployed.\\
All Transfer Lines & & \\
Booster & & \\
RR &  & \\
MI & & \\
External Beamlines (including LBNF) & &\\
Muon Campus & & 
\end{longtblr}

\begin{longtblr}[ 
        caption={Beam Control: General automations. Used stand-alone, or as a backend for the previous automations.}, 
        label={tab:beam_general_control}
    ]{
        hlines, vlines,
        row{1} = {bg=acornBlue, fg=white, font=\bfseries}, 
        rowhead = 1,
        cells = {valign=m,font=\footnotesize},
        column{1} = {0.8in,l},
        column{2} = {1.3in,l},
        column{3} = {3.7in,l}
    }
\textbf{Automations} & \textbf{Acquisition Resolution} & \textbf{Description}\\*
Smart Optimization & \SetCell[r=2]{l} upon user request or automated trigger, frequency subject to settings and instrumentation slews & Configurable in-situ optimizer and optional interface where user chooses / defines optimization algorithm, constraints, variables, and target metrics. Interfaces with Smart Scanning. Centrally deployed.\\*

Smart Scanning & & Saves state of settable variables before a scanning episode begins. Saves episodes of scans. Saves scan results and metadata in standard database. Restores state of settable variables at the end. Interfaces with Smart Optimization. Centrally deployed.\\*

Smart Timeline Generator (part of PIP-II project) & per configuration change (roughly once per day) & Expand capabilities of the TLG: Allow scheduling of different timelines (e.g. timeline A plays out 100 times, then timeline B plays. Repeat.). Fully capture timeline definition and timestamps. Enable relative indexing of repeated events in a timeline. Centrally deployed.\\*

Smart Beam Power Slew & user-defined step to ramp each consecutive super cycle & Some physical closed systems, like radioactive water cooling systems for targets or vacuum systems will have issues adapting to fast beam power increases. In these systems, it's safer to slowly ramp beam power (modify turns in Booster or rep rate in timeline). Centrally deployed.
\end{longtblr}

\subsubsection{Automations in Support System Diagnostics}
Support system diagnostics comprise hardware and software for monitoring particle accelerator infrastructure such as RF, liquid cooling, air cooling, vacuum, power, compute systems, target/horn, and consumables.
Monitoring diagnostics include new sensors tied into the control system, anomaly detection, and predictive maintenance.
\begin{itemize}
    \item Table~\ref{tab:support_sensors}: Automated sensors
    \item Table~\ref{tab:support_anomaly}: State-based anomaly monitoring
    \item Table~\ref{tab:support_predictive_maintenance}: Predictive maintenance
\end{itemize}

\begin{longtblr}[ 
        caption={Support System Diagnostics: Automated sensors.}, 
        label={tab:support_sensors}
    ]{
        hlines, vlines,
        row{1} = {bg=acornBlue, fg=white, font=\bfseries}, 
        rowhead = 1,
        cells = {valign=m,font=\footnotesize},
        column{1} = {0.8in,l},
        column{2} = {0.8in,l},
        column{3} = {4.2in,l}
    }
\textbf{Field Diagnostic} & \textbf{Acquisition Resolution} & \textbf{Description}\\*
Microphonics (\& vibration) & \SetCell[r=2]{l} continuously report as soon as possible & Use microphones, vibrational detectors (piezoelectric sensors) throughout tunnels near magnets, horns, and in hardware galleries near power supplies, pumps, and fans to automatically monitor for abnormal sounds. Only when an abnormality is found should it send data back to the control system. (Can group together by application. Compare across similar types.) \textbf{Deployed in field}; configurable centrally.\\*

Thermometers & & Use thermal sensors  throughout tunnels near magnets, horns, and in hardware galleries near power supplies, pumps, and fans to automatically monitor for abnormal tempratures. Only when an abnormality is found should it send data back to the control system. (Can group together by application. Compare across similar types.) \textbf{Deployed in field}; configurable centrally.\\

General leak detector & \SetCell[r=2]{l} continuously report as soon as possible &In service buildings and enclosures monitor sump-pumps, moisture sensors, and potentially cameras near liquid cooling systems (water) and isolation systems (fluorinert oil) and where there are anticipated building leaks. Only when an abnormality is found should it send data back to the control system. Rope moisture sensors. \textbf{Deployed in field}; configurable centrally.\\*

RF water leak detector & & Modernize our current RF low conductivity water leak detector. Only when an abnormality is found should it send data back to the control system. \textbf{Deployed in field}; configurable centrally.
\end{longtblr}

\begin{longtblr}[ 
        caption={Support System Diagnostics: State-based anomaly monitoring.}, 
        label={tab:support_anomaly}
    ]{
        hlines, vlines,
        row{1} = {bg=acornBlue, fg=white, font=\bfseries}, 
        rowhead = 1,
        cells = {valign=m,font=\footnotesize},
        column{1} = {0.9in,l},
        column{2} = {0.75in,l},
        column{3} = {4.15in,l}
    }
\textbf{System} & \textbf{Acquisition Resolution} & \textbf{Description}\\*
Cooling & \SetCell[r=8]{l} continual monitoring based on system monitoring rate & Statistical monitoring of pumps, valves, flowmeters, power supplies. Use a State Machine to identify appropriate time-dependent alarm properties. Centrally deployed.\\*

Vacuum & & Statistical monitoring of pumps, valves, flowmeters, temperature (ion pumps, turbo bearings), power supplies, RGA readings. Use a State Machine to identify appropriate time-dependent alarm properties.  Centrally deployed.\\*

Power & & Statistical monitoring power supplies and thermal components. Use a State Machine to identify appropriate time-dependent alarm properties.  Centrally deployed.\\*

RF & & Statistical monitoring of RF Gap Envelope, phase feedback, RF curves (paraphase, anode, grid, bias, frequency) compared to their error signals. Use a State Machine to identify appropriate time-dependent alarm properties.  Centrally deployed.\\*

Compute & & Statistical monitoring of networking, bandwidth, disk, memory, CPU usage, power usage, and heartbeat. Statistical and natural language processing monitoring of system and application logs.  Centrally deployed.\\*

Instrumentation &  & Statistical monitoring of instrument response to notify of hardware failures and needed re-calibration. For example, BPMs can drift over a month or two. Centrally deployed.\\*

Target / Horn & & Statistical monitoring of targets, horns, and beam trajectory consistency checks. Use a State Machine to identify appropriate time-dependent alarm properties. Centrally deployed.\\*

Consumables & & Statistical monitoring of consumables. Use a State Machine to identify appropriate time-dependent alarm properties. Examples include: DI bottle, H bottle, condensation tank, mixed gas sources for instruments and targets. Coffee. Centrally deployed.
\end{longtblr}

\begin{longtblr}[ 
        caption={Support System Diagnostics: Predictive maintenance.}, 
        label={tab:support_predictive_maintenance}
    ]{
        hlines, vlines,
        row{1} = {bg=acornBlue, fg=white, font=\bfseries}, 
        rowhead = 1,
        cells = {valign=m,font=\footnotesize},
        column{1} = {0.8in,l},
        column{2} = {0.8in,l},
        column{3} = {4.2in,l}
    }
\textbf{System} & \textbf{Acquisition Resolution} & \textbf{Description}\\*
Cooling & \SetCell[r=6]{l} continual monitoring based on system monitoring rate & \SetCell[r=5]{l} Monitoring for known failure precursors. Centrally deployed.\\

Vacuum & & \\

Power & & \\

RF & & \\

Target / Horn & & \\

Compute & & Monitoring for upgrade and maintenance scenarios (e.g. periodic reboots).  Centrally deployed.\\

Consumables & continuous prediction to future time horizon & Monitor levels of consumables over time to predict refill/replace/empty. Examples include: DI bottle, H bottle, condensation tank, dehumidifiers, air filters, mixed gas sources for instruments and targets. Coffee.  Centrally deployed.
\end{longtblr}

\subsubsection{Automations in Support System Controls}
Support system controls automations include regulation and adjustments to valves, pumps, RF, and associated PID loops, see Table~\ref{tab:support_controls}.

\begin{longtblr}[ 
        caption={Support System Controls: Automated controls.}, 
        label={tab:support_controls}
    ]{
        hlines, vlines,
        row{1} = {bg=acornBlue, fg=white, font=\bfseries}, 
        rowhead = 1,
        cells = {valign=m,font=\footnotesize},
        column{1} = {1.2in,l},
        column{2} = {1.5in,l},
        column{3} = {3.1in,l}
    }
\textbf{Automations} & \textbf{Acquisition Resolution} & \textbf{Description}\\*
Target hydrogen-argon regulation & continuous prediction to future time horizon & Monitor H-Ar ratio in target hall for LBNF to periodically vent. Centrally deployed.\\

PID autoset & upon user request (start-up, commissioning) & Create self-tuning and self-optimizing PID algorithm so that PIDs can set themselves. Also anomaly detection can help to control PID swings. Centrally deployed.\\

Further automation of valves and pumps & continual adjustments based on system monitoring rate and response rate & Change valve position / pump speed to maintain user-defined readback (e.g. LCW three-way valve, flow, etc.) May use PID autoset solution. Centrally deployed.\\

RF Feedbacks & upon user request ($<$ each 4 hours) & Booster/MI Bias supply card current adjustments. Centrally deployed.
\end{longtblr}

\section{Workflow Integration With LLMs}
\label{sec:Workflow_integration_with_llms}
As accelerator control systems grow in complexity, Large Language 
Models (LLMs) offer a practical way to streamline development, 
enhance usability, and accelerate problem-solving. LLMs can interpret 
and generate text, enabling them to search technical documentation, 
assist with code generation, and help automate routine tasks. These 
models can also analyze operational data, support troubleshooting, 
and assist with decision-making. Their ability to interface with 
control system components and adapt to evolving workflows makes them 
a useful addition for improving the efficiency of accelerator 
operations.

\subsection{LLM Use Cases}

\subsubsection{Information Retrieval}
LLMs provide natural-language access to documentation, historical 
data, and internal procedures. Rather than relying on exact keywords 
or rigid query structures, users can interact with the system 
conversationally, making information retrieval more intuitive and 
efficient.
\begin{itemize}
    \item Logs and shift summaries become searchable in narrative 
    form.
    \item Engineering notes and design documents can be summarized 
    or synthesized.
    \item Procedures can be queried interactively, supporting 
    real-time decision-making and onboarding.
    \item Paper documentation must be scanned and digitized before 
    it can be made available to an LLM.
\end{itemize}

\subsubsection{Code \& Software Development}
LLMs assist in writing, reviewing, and understanding code across 
multiple system layers. They help reduce repetitive code, enforce 
standardized interfaces, and generate practical examples that help 
new users get started.
\begin{itemize}
    \item Developers can scaffold new components quickly using 
    conversational prompts.
    \item Existing codebases can be explored and modified with LLM 
    assistance, improving maintainability.
    \item Integration tasks, such as protocol bridging or test 
    scaffolding, can be semi-automated.
    \item Embedded within editors or notebooks, LLMs provide 
    immediate context-aware support.
\end{itemize}

\subsubsection{AI Agent Roles}
LLMs can serve as lightweight assistants across both technical and administrative domains.
\begin{itemize}
    \item In operational settings, they can answer natural-language questions about system behavior, recent changes, or alerts.
    \item During development, they can recommend code completions, suggest configuration changes, or assist with version control tasks.
    \item In administrative contexts, they can help draft documentation, summarize training content, or navigate institutional procedures.
\end{itemize}

\subsection{LLM Risk Mitigation}
While LLMs offer significant benefits, their integration into 
accelerator workflows requires deliberate safeguards.
\begin{itemize}
    \item All LLM outputs should be treated as suggestions requiring 
    human oversight and testing. Users should be trained in 
    appropriate usage.
    \item Sensitive data should not be shared with external APIs. 
    Self-hosted or sandboxed models can be used to maintain privacy 
    and control.
    \item Prompt design and context management are critical for 
    reliability. Providing models with domain-relevant material 
    improves accuracy and reduces incorrect responses.
    \item All LLM-generated code or documentation should be placed 
    under version control and included in peer review processes.
    \item Since models can reflect bias or misunderstand 
    domain-specific logic, regular evaluation with 
    Fermilab-relevant content helps align their behavior with 
    internal standards.
\end{itemize}

\subsection{LLM Summary}
LLMs represent a new interface layer between humans and complex 
systems. When integrated thoughtfully, they reduce friction in both 
development and operations, enabling teams to work more efficiently 
and manage complexity with greater ease. By embedding LLMs as 
supportive tools within daily workflows, rather than relying on them 
as autonomous agents, teams can improve productivity while preserving 
reliability and accountability.

\section{Conclusions and Next Steps}
\label{sec:Conclusions}
Community workshops and stakeholder interviews have defined a vision 
for a modernized accelerator control system that supports 
AI/ML-enhanced operations. This vision emphasizes the capability for 
accelerator physicists, machine experts, and developers to access 
reliable, high-quality, and standardized data in order to build and 
deploy advanced automations. The ACORN project provides the vehicle 
to realize this vision by delivering the enabling infrastructure.

This document has outlined the requirements and technical foundations 
for an AI-ready control system. Central to this effort is data 
standardization. Establishing consistent formats, reliable 
timestamps, and interoperable data pathways ensures that accelerator 
data can be shared, validated, and used effectively for AI/ML 
applications. By combining data standardization with lifecycle 
management practices and reproducible tooling, ACORN establishes the 
groundwork for advanced automation without itself developing AI/ML 
algorithms. These measures align with broader community priorities, 
such as ensuring accelerator data is interoperable and usable in 
emerging platforms like the American Science Cloud. In doing so, ACORN 
will provide the foundation for a reliable, 
scalable, and AI-ready control system that sustains Fermilab's role 
as the nation's flagship accelerator complex. 

\clearpage

\clearpage
\section{References}
\printbibliography[heading=none]
\clearpage

\end{document}